\RequirePackage{fix-cm} % Fix LaTeX2e bugs.
\documentclass[a4paper, twoside, reqno, 12pt]{amsart}
\usepackage{fixltx2e}   % Fix LaTeX2e bugs.

\usepackage[latin1]{inputenc}
\usepackage[T1]{fontenc}

\usepackage{eucal}
\usepackage{esint}
\usepackage{dsfont}
\usepackage{xspace}
\usepackage{amsgen}
\usepackage{amsthm}
\usepackage{amssymb}
\usepackage{amsmath}
\usepackage{upgreek}
\usepackage{amsfonts}
\usepackage{mathrsfs}
\usepackage{mathtools}
\usepackage[nice]{nicefrac}

\usepackage{a4wide}

\usepackage[dvipsnames, table]{xcolor}

\definecolor{oneblue}{rgb}{0.0, 0.0, 0.85}
\definecolor{bluepigment}{rgb}{0.2, 0.2, 0.6}
\definecolor{darkgrey}{rgb}{0.273, 0.281, 0.30}
\definecolor{Lightgray}{rgb}{0.89, 0.89, 0.89}
\definecolor{Lightblue}{RGB}{214, 214, 214}
\definecolor{bckg}{RGB}{20.8, 20.8, 20.8} % Color of the boxes
\definecolor{charcoal}{rgb}{0.21, 0.27, 0.31}
\definecolor{darkelectricblue}{rgb}{0.33, 0.41, 0.47}

\usepackage[sort&compress, comma, square, numbers]{natbib}

\usepackage{psfrag}
\usepackage{graphicx}
\usepackage{subfigure}
\usepackage{morefloats}
\usepackage{indentfirst}

\usepackage{acronym}
\usepackage{microtype}
\usepackage[labelsep=period,%
            labelfont={bf,sf,color=bluepigment},%
            justification=raggedright]{caption}

\usepackage[usenames, dvipsnames]{pstricks}
\usepackage{epsfig}
\usepackage{pst-grad} % For gradients
\usepackage{pst-plot} % For axes

\usepackage{epstopdf}

\usepackage[colorlinks,
           urlcolor=oneblue,
           linkcolor=oneblue,
           citecolor=NavyBlue,
           bookmarksopen=false,
           pdfpagemode=UseNone,
           pagebackref]{hyperref}

\usepackage[explicit]{titlesec}

\titleformat{\section}
  {\color{NavyBlue}\Large\sffamily\bfseries}
  {}
  {0em}
  {\colorbox{bckg!5}{\parbox{\dimexpr\linewidth-2\fboxsep\relax}{\centering\thesection. #1}}}
  [\vspace*{0.33em}]

\titleformat{name=\section,numberless}
  {\color{NavyBlue}\Large\sffamily\bfseries}
  {}
  {0.0em}
  {\colorbox{bckg!10}{\parbox{\dimexpr\linewidth-2\fboxsep\relax}{\centering#1}}}
  [\vspace*{0.33em}]

\titleformat{\subsection}
  {\color{NavyBlue}\large\sffamily\bfseries}
  {}
  {0.0em}
  {\colorbox{bckg!5}{\parbox{\dimexpr\linewidth-2\fboxsep\relax}{\centering\thesubsection. #1}}}
  [\vspace*{0.33em}]

\titleformat{name=\subsection,numberless}
  {\color{NavyBlue}\Large\sffamily\bfseries}
  {}
  {0em}
  {\colorbox{bckg!10}{\parbox{\dimexpr\linewidth-2\fboxsep\relax}{\centering#1}}}
  [\vspace*{0.33em}]

\titleformat{\subsubsection}
  {\color{bluepigment}\sffamily\normalsize\bfseries}
  {\thesubsubsection}
  {0.5em}
  {#1}
  [\vspace*{0.33em}]

\titleformat{\paragraph}[runin]
  {\color{bluepigment}\sffamily\small\bfseries}
  {}
  {0em}
  {#1}

\titlespacing{\section}{1.0em}{1.5em plus 2pt minus 2pt}%
{1.0em plus 2pt minus 2pt}[0em]
\titlespacing{\subsection}{1.0em}{1.5em plus 2pt minus 2pt}%
{1.0em}[0em]
\titlespacing{\subsubsection}{1.0em}{1.5em plus 2pt minus 2pt}%
{1.0em plus 2pt minus 2pt}[0em]

\usepackage{titletoc}

\contentsmargin{0.5em}
\newlength{\tocsep} 
\titlecontents{section}[\tocsep]
  {\addvspace{10pt}\bfseries\sffamily}
  {\contentslabel[\thecontentslabel]{\tocsep}}
  {}
  {\ \titlerule*[0.75pc]{.}\ \thecontentspage}
  []
\titlecontents{subsection}[\tocsep]
  {\addvspace{8pt}\sffamily}
  {\contentslabel[\thecontentslabel]{\tocsep}}
  {}
  {\ \titlerule*[0.5pc]{.}\ \thecontentspage}
  []
\titlecontents*{subsubsection}[\tocsep]
  {\addvspace{2pt}\footnotesize\sffamily}
  {}
  {}
  {\ \titlerule*[0.35pc]{.}\ \thecontentspage}
  [\\*]

\makeatletter
\def\@setauthors{%
  \begingroup
  \def\thanks{\protect\thanks@warning}%
  \trivlist
  \centering\footnotesize \@topsep30\p@\relax
  \advance\@topsep by -\baselineskip
  \item\relax
  \author@andify\authors
  \def\\{\protect\linebreak}%
  \textsc{\normalsize\textcolor{darkelectricblue}{\authors}}%
  \ifx\@empty\contribs
  \else
    ,\penalty-3 \space \@setcontribs
    \@closetoccontribs
  \fi
  \endtrivlist
  \endgroup
}
\def\@settitle{\begin{center}%
  \baselineskip14\p@\relax
    \bfseries
    \textsc{\Large\textcolor{charcoal}{\@title}}
  \end{center}%
}
\makeatother

\usepackage{enumitem}
\setlist[description]{%
  topsep=30pt,               % space before start / after end of list
  itemsep=5pt,               % space between items
  font={\bfseries\sffamily\color{NavyBlue}}, % if colour is needed
}

\usepackage{fancyhdr}
\usepackage{lastpage}

\newcommand*\Title{\textcolor{bluepigment}{New asymptotic heat transfer model}}
\newcommand*\Authors{\textcolor{bluepigment}{M.~Chhay, D.~Dutykh, \etal}}
\newcommand*{\plogo}{\textcolor{gray}{{\texttt{arXiv.org} / \textsc{hal}}}} % Generic publisher logo

\numberwithin{equation}{section}

\newcommand{\0}{^{(0)}}
\newcommand{\1}{^{(1)}}

\newcommand{\bt}{\theta}

\newcommand{\R}{\mathds{R}}

\newcommand{\ud}{\mathrm{d}}

\newcommand{\eps}{\varepsilon}
\renewcommand{\O}{\mathcal{O}}

\newcommand{\bu}{\boldsymbol{u}}

\renewcommand{\gamma}{\boldsymbol{\upgamma}}

\newcommand{\B}{\mathcal{B}}
\newcommand{\MA}{\mathrm{Ma}}
\newcommand{\RE}{\mathrm{Re}}

\newcommand{\PE}{\mathrm{Pe}}
\newcommand{\WE}{\mathrm{We}}
\newcommand{\BI}{\mathrm{Bi}}
\newcommand{\PR}{\mathrm{Pr}}
\newcommand{\KA}{\mathrm{Ka}}
\newcommand{\bT}{\overline{T}}
\newcommand{\w}{\mathrm{wall}}
\newcommand{\bn}{\boldsymbol{n}}
\newcommand{\BITILDE}{\mathrm{\widetilde{Bi}}}

\newcommand{\ie}{\emph{i.e.}~}
\newcommand{\eg}{\emph{e.g.}~}

\newcommand{\etal}{\emph{et al.}~}

\renewcommand{\div}{\grad\scal}

\newcommand{\scal}{\boldsymbol{\cdot}}
\newcommand{\grad}{\boldsymbol{\nabla}}

\newcommand{\pd}[2]{\frac{\partial #1}{\partial\/ #2}}

\newcommand{\eqdef}{\mathop{\stackrel{\,\mathrm{def}}{:=}\,}}

\begin{document}

\title[\Title]{New asymptotic heat transfer model in thin liquid films}

\author[M.~Chhay]{Marx Chhay$^*$}
\address{LOCIE, UMR 5271 CNRS, Universit\'e Savoie Mont Blanc, Campus Scientifique, 73376 Le Bourget-du-Lac Cedex, France}
\email{Marx.Chhay@univ-savoie.fr}
\urladdr{http://marx.chhay.free.fr/}
\thanks{$^*$ Corresponding author}

\author[D.~Dutykh]{Denys Dutykh}
\address{LAMA, UMR 5127 CNRS, Universit\'e Savoie Mont Blanc, Campus Scientifique, 73376 Le Bourget-du-Lac Cedex, France}
\email{Denys.Dutykh@univ-savoie.fr}
\urladdr{http://www.denys-dutykh.com/}

\author[M.~Gisclon]{Marguerite Gisclon}
\address{LAMA, UMR 5127 CNRS, Universit\'e Savoie Mont Blanc, Campus Scientifique, 73376 Le Bourget-du-Lac Cedex, France}
\email{Marguerite.Gisclon@univ-savoie.fr}
\urladdr{https://www.lama.univ-savoie.fr/~gisclon/}

\author[Ch.~Ruyer-Quil]{Christian Ruyer-Quil}
\address{LOCIE, UMR 5271 CNRS, Universit\'e Savoie Mont Blanc, Campus Scientifique, 73376 Le Bourget-du-Lac Cedex, France}
\email{Christian.Ruyer-Quil@univ-savoie.fr}
\urladdr{https://www.lama.univ-savoie.fr/~gisclon/}

%%% ------------------------------------------------------------------------ %%%

\begin{titlepage}
\thispagestyle{empty} % Remove page numbering on this page
\noindent
{\Large Marx \textsc{Chhay}}\\
{\it\textcolor{gray}{LOCIE, Polytech Annecy--Chamb\'ery, France}}
\\[0.02\textheight]
{\Large Denys \textsc{Dutykh}}\\
{\it\textcolor{gray}{CNRS, Universit\'e Savoie Mont Blanc, France}}
\\[0.02\textheight]
{\Large Marguerite \textsc{Gisclon}}\\
{\it\textcolor{gray}{CNRS, Universit\'e Savoie Mont Blanc, France}}
\\[0.02\textheight]
{\Large Christian \textsc{Ruyer-Quil}}\\
{\it\textcolor{gray}{LOCIE, Polytech Annecy--Chamb\'ery, France}}
\\[0.16\textheight]

\colorbox{Lightblue}{
  \parbox[t]{1.0\textwidth}{
    \centering\huge\sc
    \vspace*{0.79cm}
    
    \textcolor{bluepigment}{New asymptotic heat transfer model in thin liquid films}
    
    \vspace*{0.79cm}
  }
}

\vfill % Whitespace between the title block and the publisher

\raggedleft     % Right-align all text
{\large \plogo} % Publisher and logo
\end{titlepage}

%%% ------------------------------------------------------------------------ %%%

\newpage
\maketitle
\thispagestyle{empty}

%%% ------------------------------------------------------------------------ %%%

\begin{abstract}

In this article, we present a model of heat transfer occurring through a li\-quid film flowing down a vertical wall. This new model is formally derived using the method of asymptotic expansions by introducing appropriately chosen dimensionless variables. In our study the small parameter, known as the film parameter, is chosen as the ratio of the flow depth to the characteristic wavelength. A new \textsc{Nusselt} solution should be explained, taking into account the hydrodynamic free surface variations and the contributions of the higher order terms coming from temperature variation effects. Comparisons are made with numerical solutions of the full \textsc{Fourier} equations in a steady state frame. The flow and heat transfer are coupled through \textsc{Marangoni} and temperature dependent viscosity effects. Even if these effects have been considered separately before, here a fully coupled model is proposed. Another novelty consists in the asymptotic approach in contrast to the weighted residual approach which have been formerly applied to these problems.

\bigskip
\noindent \textbf{\keywordsname:} Heat transfer; Thin liquid film; Asymptotic modelling; Long waves; Thermal dependency properties; Marangoni effect \\

\smallskip
\noindent \textbf{MSC:} \subjclass[2010]{76D33 (primary), 76B25, 76B15 (secondary)}

\end{abstract}

%%% ------------------------------------------------------------------------ %%%

\newpage
\tableofcontents
\thispagestyle{empty}

%%% ------------------------------------------------------------------------ %%%

\newpage
\section{Introduction}

Liquid films have many significant applications in chemical engineering because of their reduced resistance to heat and mass transfers. This thermal resistance may be further reduced by inducing surface deformations and wave patterns as surface instabilities may be triggered by the dependence of surface tension to temperature (\textsc{Marangoni} effect \cite{Joo1991}) or by an hydrodynamic mechanism when the fluid is set into motion either by gravity (\textsc{Kapitza} instability mode \cite{Kalliadasis2012}) or by centrifugal acceleration \cite{Matar2005}. As an example, the \textsc{Marangoni} effect may be coupled to wall topography to induce thermocapillary convection \cite{Alexeev2005} and to promote heat transfer.

In this paper, we focus on falling liquid films. These flows are generally encountered whenever the pressure drop is critical, \eg in absorption machines, or whenever a low thermal driving force is required, for instance in the separation of multicomponent mixtures that are temperature-dependent. The dynamics of such flows have attracted a considerable interest as it presents a wavy regime organized around large-amplitude tear-drop like solitary waves whose interactions intensify transfers. This wavy regime is triggered by a long-wave instability mode corresponding to a zero critical wavenumber. For this reason, the waves are long compared to the film thickness, they emerge at relatively long distances from the liquid inlet and they are slow to interact one with another. As a result, direct numerical simulations (DNSs) of such flows are hindered by the large domain that is necessary to account for their natural evolution, which explains that DNSs are generally restricted to two-dimensional, \ie spanwise independent, situations or to the construction of periodic waves.

Mathematical modeling offers a useful reduction of the numerical cost and a welcome framework for the understanding of the disordered dynamics of such flows with the development of coherent-structure theories. Indeed, the large aspect ratio of the waves enables to introduce a small parameter $\eps$, or film parameter, which compares the typical length of the wave to the thickness of the film. In this framework, the streamwise ($x$) and spanwise ($y$) coordinates as well as the time ($t$) are slow variables, \ie $\partial_{\,x,\,y,\,t}\ \propto\ \eps\,$, whereas the cross-stream coordinate is a fast variable ($\partial_{\,z}\ =\ \O(1)$). It is thus possible to eliminate the fast variable $z$ and to obtain a reduced set of equations which describes the slow evolution of the film in a spatial domain whose dimension is reduced from 3D to 2D or from 2D to 1D if spanwise independent solutions are looked after. Following \textsc{Kapitza}'s initial work \cite{Kapitza1948}, an important amount of work has been produced in order to derive such reduced set of equations or low-dimensional models (see for instance the review by \cite{Kalliadasis2012}). \textsc{Benney} \cite{Benney1966} thus showed that a series expansion of the flow variables with respect to the film parameter $\eps$ leads to a solution that is fully characterized by the film thickness $h$ and its gradients, the film dynamics being governed by a single evolution equation for $h\,$. Unfortunately, \textsc{Benney}'s equation admits non-physical singularities in finite time at moderate \textsc{Reynolds} number \cite{Pumir1983} as a result of a too strict slaving of the velocity field to the free surface elevation. A cure to this shortcoming is offered within the Saint-Venant framework after averaging the primitive equation across the film depth. This idea dates back to the original work of \textsc{Kapitza} \cite{Kapitza1948} and was successfully applied by \textsc{Shkadov} \cite{Shkadov1967} who derived a set of two evolution equations for the local thickness $h$ and the local flow rate $q\,$. \textsc{Shkadov}'s averaging approach requires a closure hypothesis in the form of a polynomial ansatz for the velocity distribution, which corresponds to the \textsc{Nusselt} parabolic profile in \textsc{Shkadov}'s classical work. More sophisticated ansatz have been proposed \eg in \cite{Yu1995, Bohr1997}. However, consistent averaging of the primitive equations has been introduced by \textsc{Roberts} \cite{Roberts1996} and \textsc{Ruyer-Quil} \& \textsc{Manneville} \cite{Ruyer-Quil2000} using different approaches. Currently, one of the widely used approaches is the \emph{Weighted Residual Method} (WRM), which has been successfully applied to derive approximate equations \cite{Ruyer-Quil2000, Kalliadasis2012}. However, the main drawback of the WRM is that the mathematical structure of the resulting averaged equations is unclear. Consequently, the models derived in this way may be difficult to justify mathematically. This is the main reason why we opt for the method of asymptotic expansions in the present study. Other references on this topic include \cite{Luchini2010, Rojas2010, Benilov2014}. Note that the approach by \textsc{Luchini} and \textsc{Charru} \cite{Luchini2010}, based on the averaging of the kinetic energy equation of the flow, leads to a result that is identical to the WRM method at first-order of the film parameter. In essence, the \emph{Inertial Lubrication Theory} introduced by \cite{Rojas2010}, is similar to the WRM method, leading to very similar equations, coefficients being almost identical for most terms. \textsc{Benilov} \cite{Benilov2014} derived depth-averaged equations using the \textsc{Galerkin} method, which is one particular weighted residual method.

\textsc{Benney}'s original work has been extended in \cite{Joo1991, Joo1991a} to deal with the conduction of heat across the film and the coupling of the hydrodynamics to the transfer offered by the dependence of surface tension on temperature (\textsc{Marangoni} effects). The classical \textsc{Benney} expansion followed for instance by \cite{Gambaryan-Roisman2010, Leslie2011, Leslie2012} to deal with the heat transfer across an horizontal film is not appropriate to deal with a falling film. Indeed, this approach requires that heat diffusion overcomes convection, and is thus limited to only small values of the \textsc{P\'eclet} number. To overcome this limitation, \textsc{Scheid} \etal followed the Weighted Residual technique initiated by \textsc{Ruyer-Quil} \& \textsc{Manneville} \cite{Ruyer-Quil2000} and derived several models of various accuracy \cite{Ruyer-Quil2005, Scheid2005}. Though enabling to accurately decipher the complex interplay between the \textsc{Kapitza} hydrodynamic instability and the long-wave \textsc{Marangoni} thermo-capillary instability \cite{Scheid2008}, these models are only valid at still relatively low values of the \textsc{P\'eclet} number. Indeed, as the \textsc{P\'eclet} number is raised, these models may predict nonphysical values of the temperature. This behaviour is related to the onset of sharp temperature gradients at the free surface due to flow orientation in large-amplitude solitary waves. Though a cure to this limitation has been proposed by \cite{Trevelyan2007}, available low-dimensional models still fail to capture correctly the temperature distribution at large \textsc{P\'eclet} number.

This article aims at deriving a new conservative formulation for heated falling film wich retains consistency with the classical long-wave expansion. This new formulation can be seen as a low-dimensional modelling of heated falling film flows following a derivation procedure that has been proposed by \textsc{Vila} and coworkers \cite{Boutounet2008, Fernandez-Nieto2010, Boutounet2013, Noble2007}. This procedure, which will be referred to hereinafter as the \textsc{Saint}-\textsc{Venant} consistent approach, is based on the classical \textsc{Saint}-\textsc{Venant} equations that are obtained by in-depth averaging of the primitive equations with a uniform weight. However, contrary to the \textsc{Kapitza}--\textsc{Shkadov} approach which assumes the velocity field to be strictly parabolic \cite{Shkadov1967}, \textsc{Vila} proposes a closure that is compatible with \textsc{Benney}'s long-wave asymptotics and enables to accurately recover the threshold of the \textsc{Kapitza} instability. Our derivation takes into account the thermocapillary effect and the dependence of viscosity with respect to temperature. These are the principal coupling mechanisms between the hydrodynamics of the film and the heat transfer induced by the dependence of the thermophysical properties of the fluid with temperature \cite{DAlessio2014, Alexeev2005, Gambaryan-Roisman2010}.

The structure of the paper is as follows. In the next Section, governing equations are recalled. Then some physical behaviour of the heated falling film are highlighted, using a basic modeling in order to introduce the dynamic of the system. In Section~\ref{asymptotic_model} the asymptotic model is derived.  In Section~\ref{sec:num}, some numerical experiments illustrate in one hand the realistic behaviour of the model and, on the other hand, a comparison with an existing model in the literature is performed. Some discussions about the formal derivation conclude this work in Section~\ref{sec:disc}.

%%% ------------------------------------------------------------------------ %%%

\section{Problem formulation}

\subsection{Governing equations}

\begin{figure}
  \centering
  \includegraphics[width=0.88\textwidth]{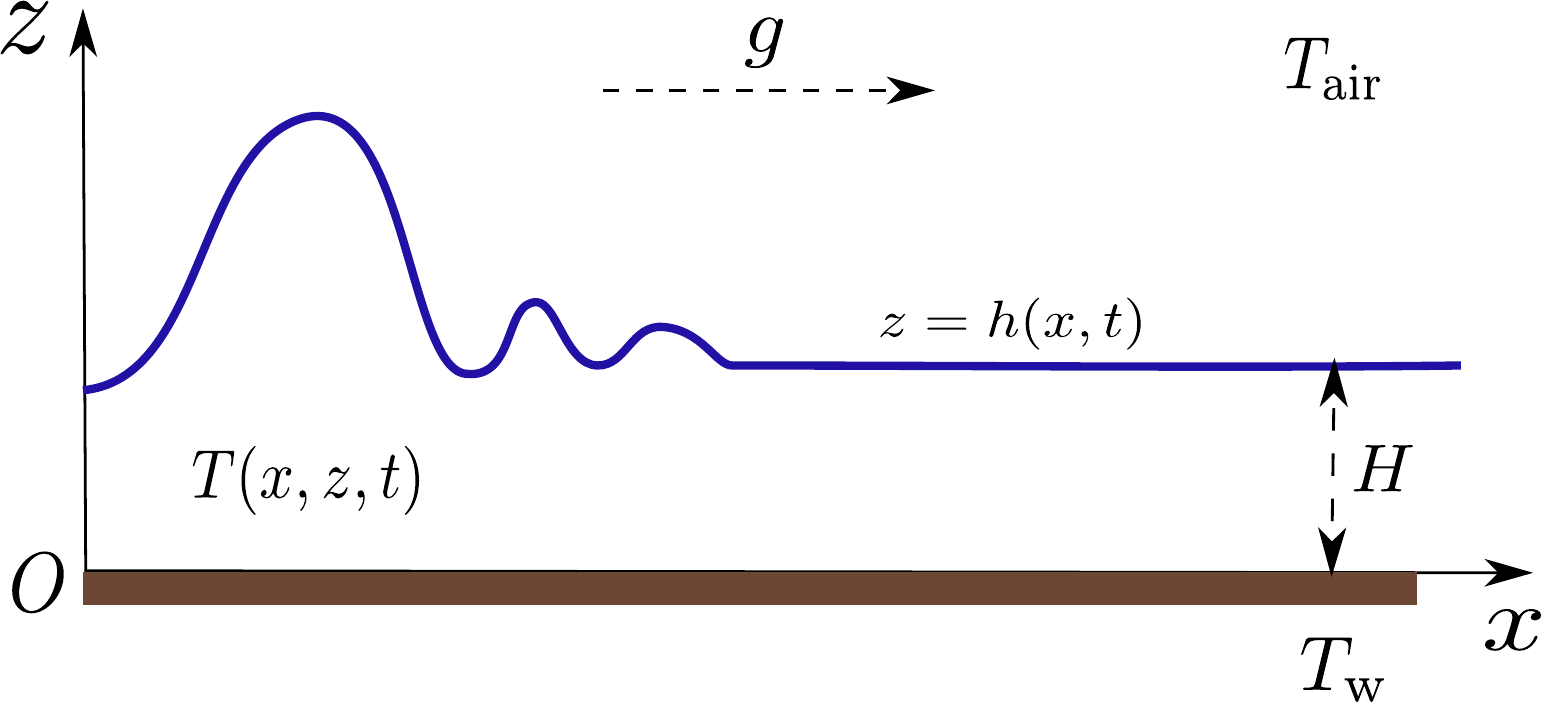}
  \caption{\small\em Sketch of the physical fluid domain.}
  \label{fig:sketch}
\end{figure}

We consider an anisotherm liquid film flowing down a heated vertical plate. The flow is supposed to be two-dimensional in space, the $x-$axis corresponding to the streamwise direction and the $z-$axis to the cross-stream direction. The liquid domain is delimited by a vertical wall at $z\ =\ 0$ and the free surface boundary located at the height $z\ =\ h(x,t)\,$. The sketch of the fluid domain is depicted in Figure~\ref{fig:sketch}.

The motion of the liquid is governed by the incompressible \textsc{Navier}--\textsc{Stokes} equations
\begin{align*}
  \rho\,\partial_t \bu\ +\ \rho \ (\bu \scal \grad ) \bu\ +\ \grad p\ -\ \rho \,  \mathbf{g}\ &=\ \div(\mu\, \grad\bu)\,, \\
  \div\bu\ &=\ 0\,,
\end{align*}
where $\bu\ =\ (u,\,w)\,$, $p$ and $\mathbf{g}$ represent the velocity and pressure fields and the gravity acceleration vector. The physical parameters $\rho\,$, $\nu\,$, $\mu$ correspond to the density, the kinematic viscosity and the dynamic viscosity. The heat transfer and the hydrodynamics of the film are coupled by the dependence of the physical properties with respect to the temperature. In this paper we focus on the principal sources of coupling and assume that only the surface tension $\sigma$ and the dynamic viscosity of the liquid $\mu\ =\ \rho \nu$ depend on the temperature $T\,$. As a further simplification, we assume linear laws, $$\mu\ =\ \mu(T)\ =\ \mu_0\ -\ \mu_1\,(T\ -\ T_0)$$ and $$\sigma\ =\ \sigma(T)\ =\ \sigma_0\ -\ m\,(T\ -\ T_0)\,$$
 where $\mu_1\ \eqdef\ -\dfrac{\ud \mu}{\ud T}$ and $m\ \eqdef\ -\dfrac{\ud \sigma}{\ud T}$ are positive constants as surface tension and viscosity generally decrease with the temperature.

The heat transfer occurring through the liquid domain is modelled by the advection-diffusion \textsc{Fourier} equation
\begin{equation*}
  \partial_t T\ +\ \bu \scal \grad T\ =\ \alpha\, \grad^2 T
\end{equation*}
where $T$ corresponds to the temperature field and $\alpha$ is the thermal diffusion coefficient, which is assumed to remain constant.

At the wall, a no-slip condition 
\begin{equation}\label{mur}
  \left.\bu\,\right|_{z = 0}\ =\ 0
\end{equation}
and a constant wall temperature $\left.T\,\right|_{z=0}\ =\ T_\w$ are imposed while at the free surface, the kinematic condition governing the evolution of the fluid elevation reads
\begin{equation}\label{hauteur}
  \partial_t\, h\ +\ \left.u\,\right|_{z\, =\, h}\cdot\partial_x h\ =\ \left.w\,\right|_{z\, =\, h}\,.
\end{equation}

The continuity of the fluid stresses at the free surface gives
\begin{align*}
  \left.p\,\right|_{z\, =\, h}\ +\ \sigma\;\frac{\partial_{xx}^2\, h}{[1\ +\ (\partial_x h)^2]^{\,3/2}}\ =\ -2\,\mu\, \frac{1\ +\ (\partial_x h)^2}{1\ -\ (\partial_xh)^2}\left.\partial_x u\,\right|_{z\, =\, h}, \\
  \dfrac{\mu}{n}\,\left[ 2\,\partial_x h\,(\partial_z w -\partial_x u)\ +\ (1 - (\partial_x h)^2)(\partial_z u + \partial_x w)\,\right]\ +\ m\,(\partial_x T + \partial_x h\,\partial_z T)\ =\ 0\,,
\end{align*}
where $n\ \eqdef\ \sqrt{1\ +\ (\partial_x h)^2}\,$.

Heat transfer at the free surface is modelled by a thermal exchange coefficient $\mathfrak{h}$ that is assumed to remain constant so that temperature field verifies the \textsc{Newton}'s law of cooling at the free surface
\begin{equation*}
  - \lambda\left.(\grad T \scal \bn)\,\right|_{z\, =\, h}\ =\ \mathfrak{h}\;\bigl[\left.T\,\right|_{z\, =\, h}\ -\ T_{\mathrm{air}}\bigr]
\end{equation*}
where $\lambda$ and $\bn$ denote the thermal conductivity and the unit exterior normal
\begin{equation*}
  \bn\ =\ \dfrac{1}{\sqrt{1\ +\ (\partial_x h)^2}}\,\left(
    \begin{array}{c}
       -\partial_x\, h \\
         1
    \end{array}\right)\,.
\end{equation*}

%%% ------------------------------------------------------------------------ %%%

\subsection{Scaled equations}

The specific geometry of the falling film is characterized by the typical length scales in both the streamwise direction and the cross-stream direction. The evolution of the hydrodynamic instabilities and the thermal diffusion process can also be described through these typical lengths. Introducing the dimensional quantities
\begin{itemize}
  \item $L$ : streamwise typical length scale,
  \item $H$ : cross-stream typical length scale,
  \item ${U_0\ \eqdef\ \dfrac{\rho\,g\,H^2}{2\,\mu_0}}$ : typical average velocity corresponding to hydrodynamic \textsc{Nusselt} solution
\end{itemize}
and the following change of variables \cite{Noble2007}:
\begin{equation*}
  t\ =\ \bar{t}\; \frac{L}{U_0}, \quad
  x\ =\ \bar{x}\; L, \quad
  z\ =\ \bar{z}\; H, \quad
  h\ =\ \bar{h}\; H, \quad
  u\ =\ \bar{u}\; U_0,
\end{equation*}
\begin{equation*}
  w\ =\ \bar{w}\; U_0 \dfrac{H}{L}, \quad
  p\ =\ \bar{p}\; \rho \, g  \, H, \quad
  T\ =\ \bar{T}\; (T_{\w}\ -\ T_{\mathrm{air}})\ +\ T_{\mathrm{air}}\,,
\end{equation*}
six dimensionless numbers characterize the problem at hand:
\begin{itemize}
  \item the \textsc{Reynolds} number $\RE\ \eqdef\ \dfrac{\rho\,U_0\,H}{\mu_0}\,$,
  \item the \textsc{P\'eclet} number $\PE\ \eqdef\ \dfrac{U_0\, H}{\alpha}\,$,
  \item the \textsc{Biot} number $\BI\ \eqdef\ \dfrac{\mathfrak{h}\,H}{\lambda}\,$,
 \item the \textsc{Weber} number $\WE\ \eqdef\ \dfrac{\sigma_0}{\rho\, g\, H^2}\,$, 
 \item the \textsc{Marangoni} number $\MA\ \eqdef\ \dfrac{2\,m\,(T_{\mathrm{wall}}\ -\ T_{\mathrm{air}})}{\rho\,g\,H^2}\ =\ \dfrac{m\,(T_{\mathrm{wall}}\ -\ T_{\mathrm{air}})}{\mu_0 \, U_0}\,$,
  \item the dimensionless \emph{rate of change} of the dynamic viscosity $$\Pi_{\mu}\ \eqdef\ \dfrac{\mu_1}{\mu_0}\;(T_{\mathrm{wall}}\ -\ T_{\mathrm{air}})\,,$$
 \end{itemize}
besides the film parameter $\eps\ \eqdef\ \dfrac{H}{L}$. In what follows, the reference temperature is further chosen to correspond to the temperature of the air, \ie $T_0\ =\ T_{\mathrm{air}}\,$. For most liquids, surface tension is high and the \textsc{Weber} number is typically large. We thus introduce $\WE\ \eqdef\ \dfrac{\kappa}{\eps^2}$ with $\kappa\ =\ \O(1)\,$. This classical assumption (see \cite{Kalliadasis2012}) enables to include surface tension effects at an early stage of the asymptotic long-wave assumption, as surface tension is the sole physical effect which prevent the breaking of the waves. The coupling of the flow dynamics with respect to the heat transfer is accounted for by the \textsc{Marangoni} number $\MA$ and the dimensionless rate of change of the viscosity $\Pi_\mu$ that are assumed to be order one quantities. Note that the positivity of viscosity and surface tension demands that $\Pi_\mu < 1$ and $\MA < 2 \WE$.

This set of parameters is usefully completed with the \textsc{Kapitza} number $\KA\ \eqdef\ (l_c\,/\,l_\nu)^2\ =\ \WE\,(H\,/\,l_\nu)^2$ and $\BITILDE\ =\ \mathfrak{h}\,l_\nu\,/\,\lambda\ =\ \BI\,l_\nu\,/\,H\,$, where $l_c\ =\ \sqrt{\kappa/\rho\, g}$ is the capillary length, and $l_\nu\ =\ (\nu^2/g)^{1/3}$ is a viscous-gravity length \cite{Kalliadasis2012}. The dimensionless groups $\KA$ and $\BITILDE$ are independent of the film thickness $H$ and depend only on the fluid properties. The thin liquid depth is characterized by the ratio $H\ \ll\ L\,$ \ie $\eps \ll 1\,$.

The dimensionless incompressible \textsc{Fourier}--\textsc{Navier}--\textsc{Stokes} equations read 
\begin{subequations}\label{FNS-non-dim}
\begin{multline}\label{equas1} 
  \eps\,\RE\,\left(\partial_t u\ +\ u \, \partial_x u\ +\ w\,\partial_z u\ +\ \dfrac{2}{\RE}\;\partial_x p\right)\ =\\ 2\ +\ (1\ -\ \Pi_{\mu}\,T) \,(\eps^2 \partial^2_{xx}u\ +\ \partial^2_{zz} u)\ -\ \Pi_{\mu}\,(\eps^2 \partial_x T\, \partial_x u\ +\ \partial_z u\, \partial_z T)\,,
\end{multline}
\begin{multline}\label{equas2} 
  \eps\,\RE\,\left(\partial_t w\ +\ u\,\partial_x w\ +\ w \, \partial_z w\ +\ \dfrac{2}{\eps^2 \RE} \,\partial_z p \right)\ =\\ (1\ -\ \Pi_\mu T)\,(\eps^2 \partial^2_{xx}w\ +\ \partial^2_{zz} w)\ -\ \Pi_{\mu}(\eps^2 \partial_x T \partial_x u\ +\ \partial_z u\,\partial_z T)
\end{multline}
\begin{align}\label{equas3}
 \partial_x u\ +\ \partial_z w\ &=\ 0\,, \\
 \partial_t T\ +\ u\,\partial_x T\ +\ w\, \partial_z T\ &=\ \dfrac{1}{\eps\,\PE}\;\left(\eps^2\,\partial_{xx}^2 T\ +\ \partial^2_{zz} T\right)\,. \label{equas4}
\end{align}
For convenience, the over bar notation for the dimensionless quantities have been dropped in the above equations. The no-slip boundary condition at the wall \eqref{mur} and the kinematic condition at the free surface \eqref{hauteur} remain formally unmodified. The continuity conditions of the fluid stress across the free surface become
\begin{multline}\label{conditionp}
  \left.p\right|_{\,h}\ + \ \eps^2\,\left(\WE\ -\ \dfrac{1}{2}\;\MA\, \left.T\,\right|_{z\,=\,h}\right)\,\frac{\partial_{xx}^2 h}{[\,1\ +\ \eps^2\,(\partial_x h)^2\,]^{\,3/2}}\ =\\
  -\eps\,\bigl(1\ -\ \Pi_{\mu}\,\left.T\,\right|_{z\,=\,h}\bigr)\,\frac{1\ +\ \eps^2\, (\partial_x h)^2}{1\ -\ \eps^2\, (\partial_x h)^2}\left.\partial_x u\,\right|_{h}
\end{multline}
and
\begin{multline}\label{Marangoni}
(1\ -\ \Pi_{\mu}\left.T\,\right|_{z\,=\,h}) \sqrt{1\ +\ \eps^2 (\partial_x h)^2}   \Bigl[\, \bigl(1\ - \eps^2(\partial_x h)^2\bigr)\cdot\bigl(\left.\partial_z u\,\right|_{h}\ +\ \eps^2\left.\partial_{x} w\,\right|_{h}\bigr)\\ 
 +\ 4\, \eps^2\, \partial_x h  \left.\partial_z w\,\right|_{h}\,\Bigr]\ +\ \eps\,\MA\,\sqrt{1\ +\ \eps^2 (\partial_x h)^2}\;\dfrac{\ud T(h)}{\ud x}\ =\ 0
\end{multline}
where the total derivative $\dfrac{\ud T(h)}{\ud x}$ is given by
\begin{equation*}
  \dfrac{\ud T(h)}{\ud x}\ \eqdef\ \left.\bigl[\,\partial_x T\ +\ \partial_x h\cdot\partial_z T\,\bigr]\,\right|_{h}\,.
\end{equation*}

The dimensionless heat transfer between the heated liquid and the am\-bient air becomes
\begin{equation}
\label{BC-heat-fs}
 \left.\partial_z T\,\right|_{h}\ =\ -\sqrt{1\ +\ (\eps\partial_x h)^2}\ \BI \left.T\,\right|_{h}\ +\ \eps^2\,\partial_x h\,\left.\partial_x T\,\right|_{h}\,, 
\end{equation}
whereas the \textsc{Dirichlet}-type boundary condition at the wall is
\begin{equation}\label{BC-heat-wall}
  \left.T\,\right|_{z\,=\,0}\ =\ 1\,.
\end{equation}
\end{subequations}

%%% ------------------------------------------------------------------------ %%%

\subsection{Fourier full 2D model}

In order to illustrate numerically the heat transfer behaviour depending on relevant physical parameters, we present below some numerical simulations of the \textsc{Fourier} equation, the solution to the \textsc{Navier}--\textsc{Stokes} equation being approximated by the low-dimensional model that is presented below.

The numerical solution is looked after in a stationary rectangular domain thanks to the change of variables
\begin{equation*}
 \psi:\ (x,\, z,\, t)\ \mapsto\ \left(x,\, y\, =\, \frac{z}{h(x,\,t)},\, t\right)\ \in\ [0,\, L] \times [0,\, 1] \times \R_+\,.
\end{equation*}
The transformed heat field $\theta\ = \ T \circ \psi^{\,-1}$ becomes a solution of
\begin{equation}\label{fourier} 
  \eps\, \PE\,\left(\mathrm{D}_{t,\,h}\,\theta\ +\ \tilde{\bu} \scal \grad \theta \right)\ =\ \Delta_h^2 \theta\,,
\end{equation}
where the differential operators are given by
\begin{eqnarray*}
  \mathrm{D}_{t,h}\,\theta\ & = &\ \partial_t \theta\ -\ \dfrac{y}{h}\,\partial_t h\,  \partial_y \theta, \\
  \nabla_h \theta\ & = &\
 \begin{pmatrix}
  \partial_x \theta\ -\ \dfrac{y}{h}\;\partial_x h  \; \partial_y \theta \\
  \dfrac{1}{h(x)}\; \partial_y \theta
 \end{pmatrix}, \\
  \Delta_h^2 \theta\ &=&\ \eps^2\; \Bigl[\,\partial_{xx}^2 \theta\ -\ 2\; \dfrac{y}{h}\;\partial_x h\, \partial_{xy}^2 \theta\ +\ \dfrac{y}{h}\; \left( \dfrac{2}{h}\;(\partial_x h)^2\ -\ \partial_{xx}^2 h \right) \partial_y\,\theta\,\Bigr] \\ && 
  +\ \Bigl[\,\eps^2 \, \left(\dfrac{y}{h} \right)^2(\partial_x h)^2\ +\ \dfrac{1}{h^2}\,\Bigr]\,\partial_{yy}^2\, \theta
\end{eqnarray*}
and $\tilde{\bu}\ =\ (u\ -\ c,\,w)$ corresponds to the velocity vector field shifted by the wave celerity $c\,$.

The boundary condition at the wall remains
\begin{equation*}
  \left.\theta\,\right|_{y\,=\,0}\ =\ 1\,,
\end{equation*}
and the \textsc{Robin}-type condition at the free surface becomes
\begin{equation*}
  \bigl(1\ +\ (\eps\,\partial_x h)^2\bigr)\left.\partial_y \theta\,\right|_{y\, =\, 1}\ -\ \eps^2\,h\,\partial_x h \,\left.\partial_x \theta\,\right|_{y\, =\, 1}\ = -h\, \sqrt{1\ +\ (\eps \, \partial_x h)^2} \ \BI \ \left.\theta\,\right|_{y\, =\, 1}\,.
\end{equation*}
The velocity field is computed using the parabolic approximation for the downstream component $u(x,\,z)\ =\ 3\,\Bigl(z\ -\ \dfrac{1}{2}\;z^2\Bigr)\, q\, h^{-1}\ -\ c$ with $q\ =\ \dfrac{1}{3} + c\,(h\ -\ 1)$ and $c\ =\ 2.96\,$. The cross-stream velocity component $w(x,z)$ is computed such that the incompressibility is verified.

\begin{figure}
  \centering
  \vspace*{-3.5em}
  \subfigure[$\BITILDE\, =\, 0.1,\ \PR\, =\, 7$]{\includegraphics[width=0.48\textwidth]{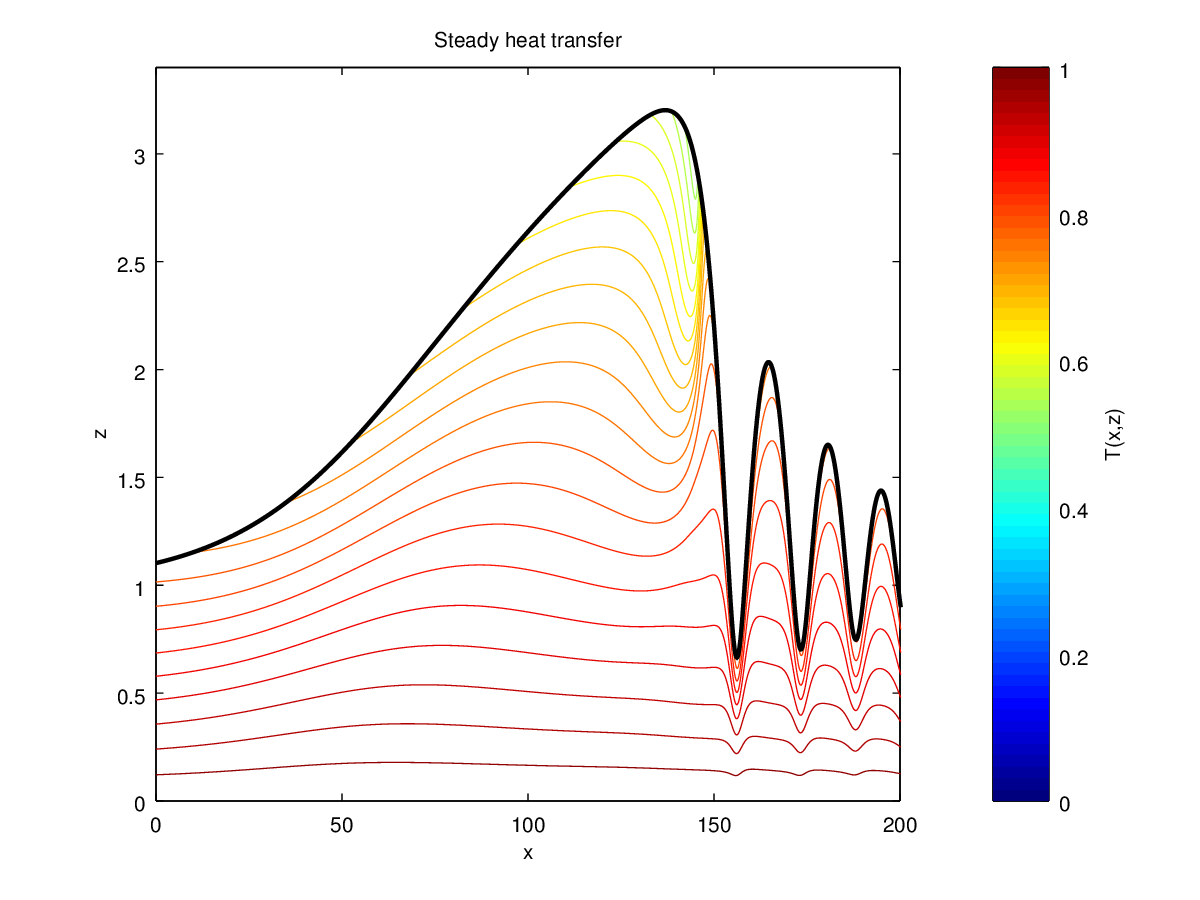}} \hfill
  \subfigure[$\PR\, =\, 1,\ \BITILDE\, =\, 0.01$]{\includegraphics[width=0.48\textwidth]{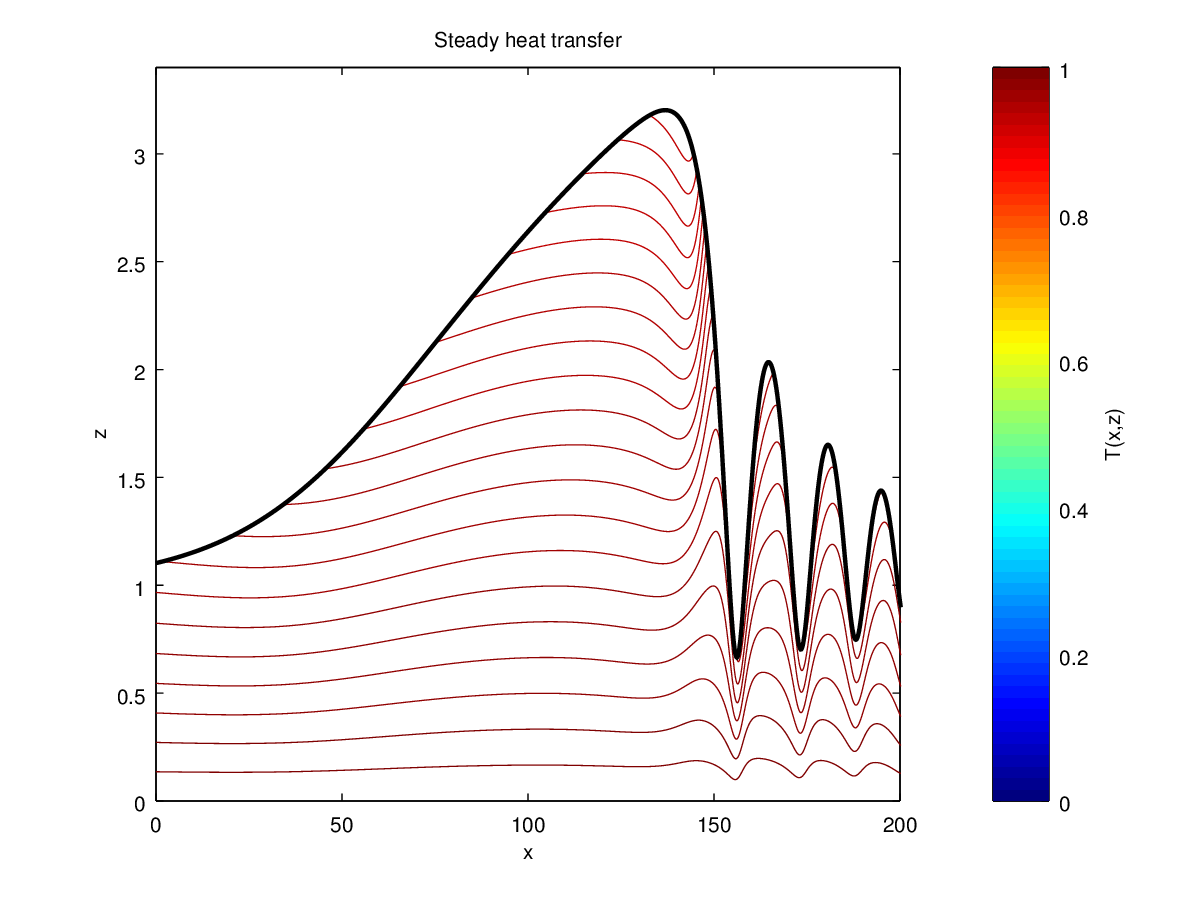}}  \hfill
  \subfigure[$\BITILDE\, =\, 1,\ \PR\, =\, 7$]{\includegraphics[width=0.48\textwidth]{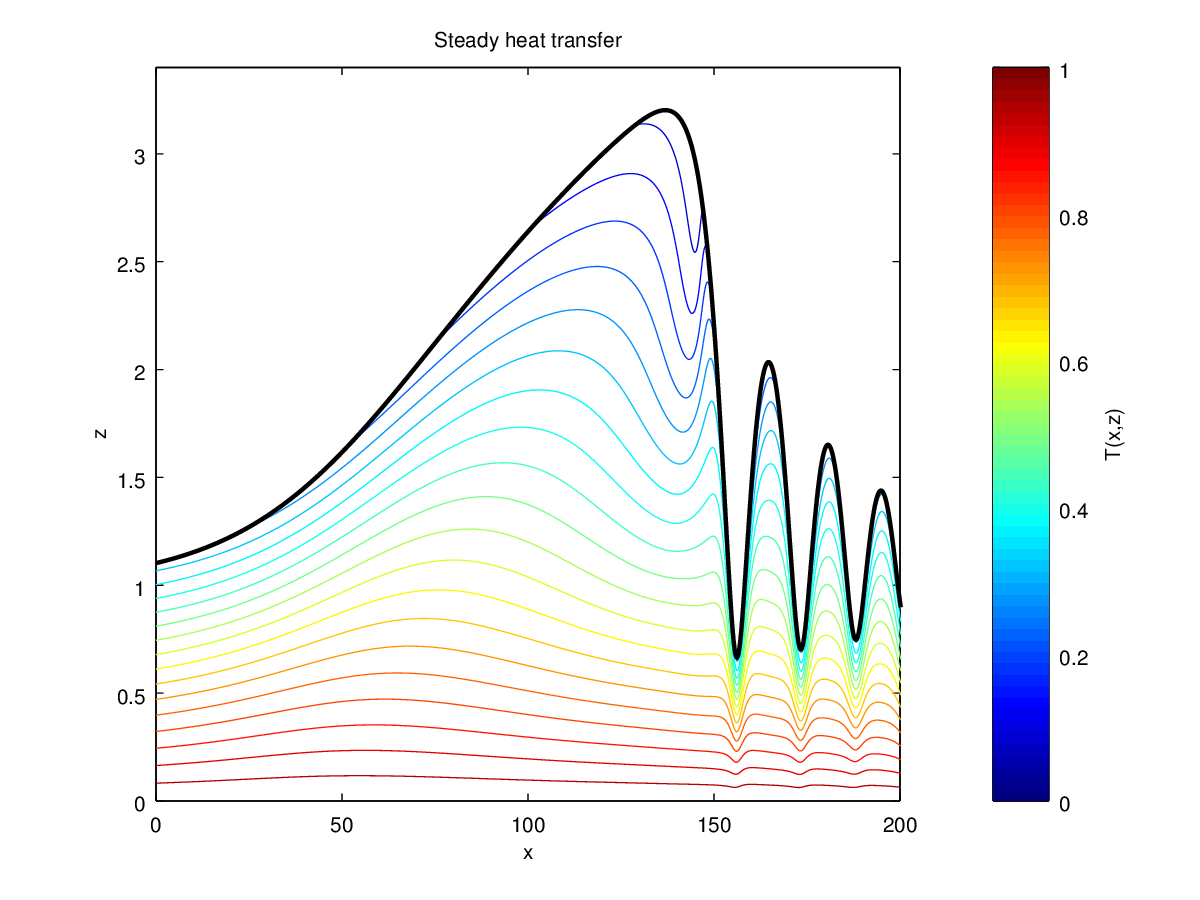}} \hfill
  \subfigure[$\PR\, =\, 7,\ \BITILDE\, =\, 0.01$]{\includegraphics[width=0.48\textwidth]{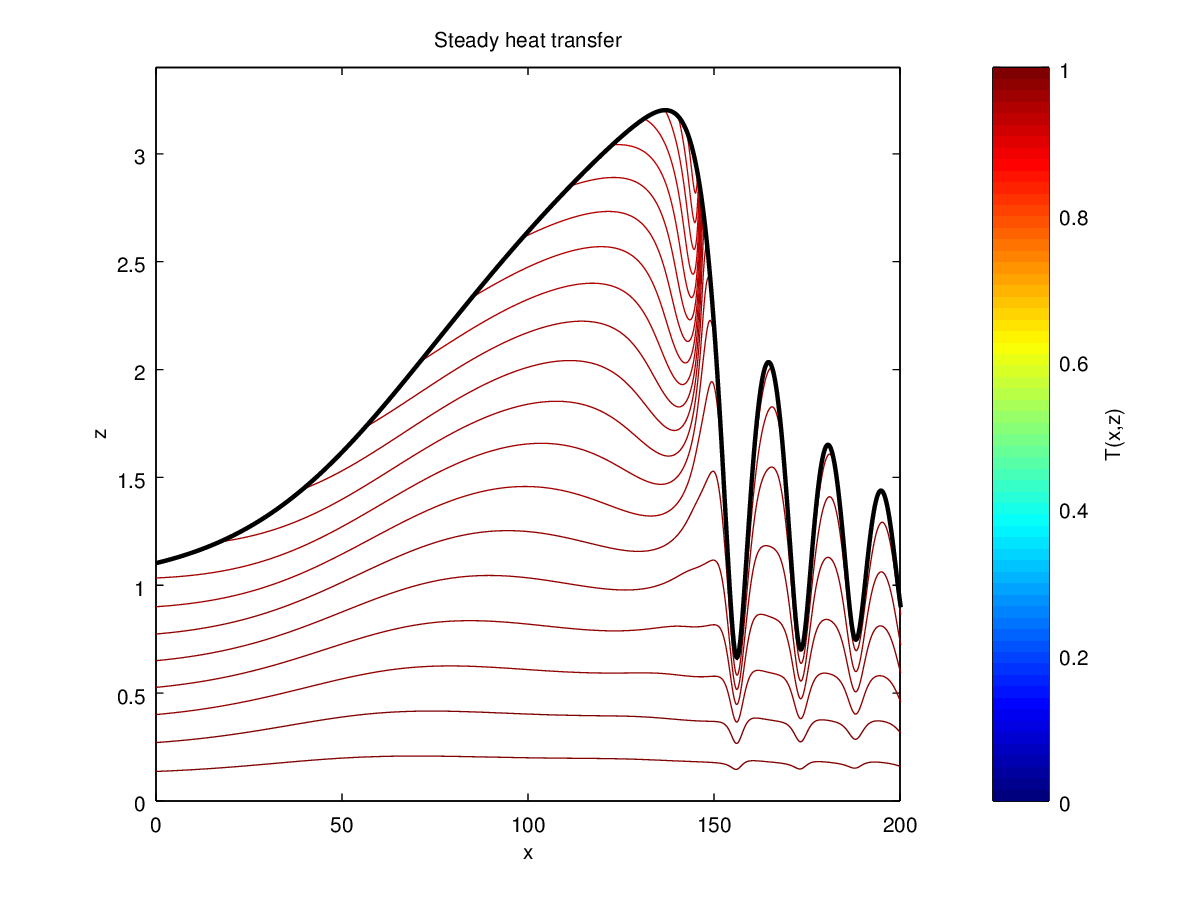}}  \hfill
  \subfigure[$\BITILDE\, =\, 10,\ \PR\, =\, 7$]{\includegraphics[width=0.48\textwidth]{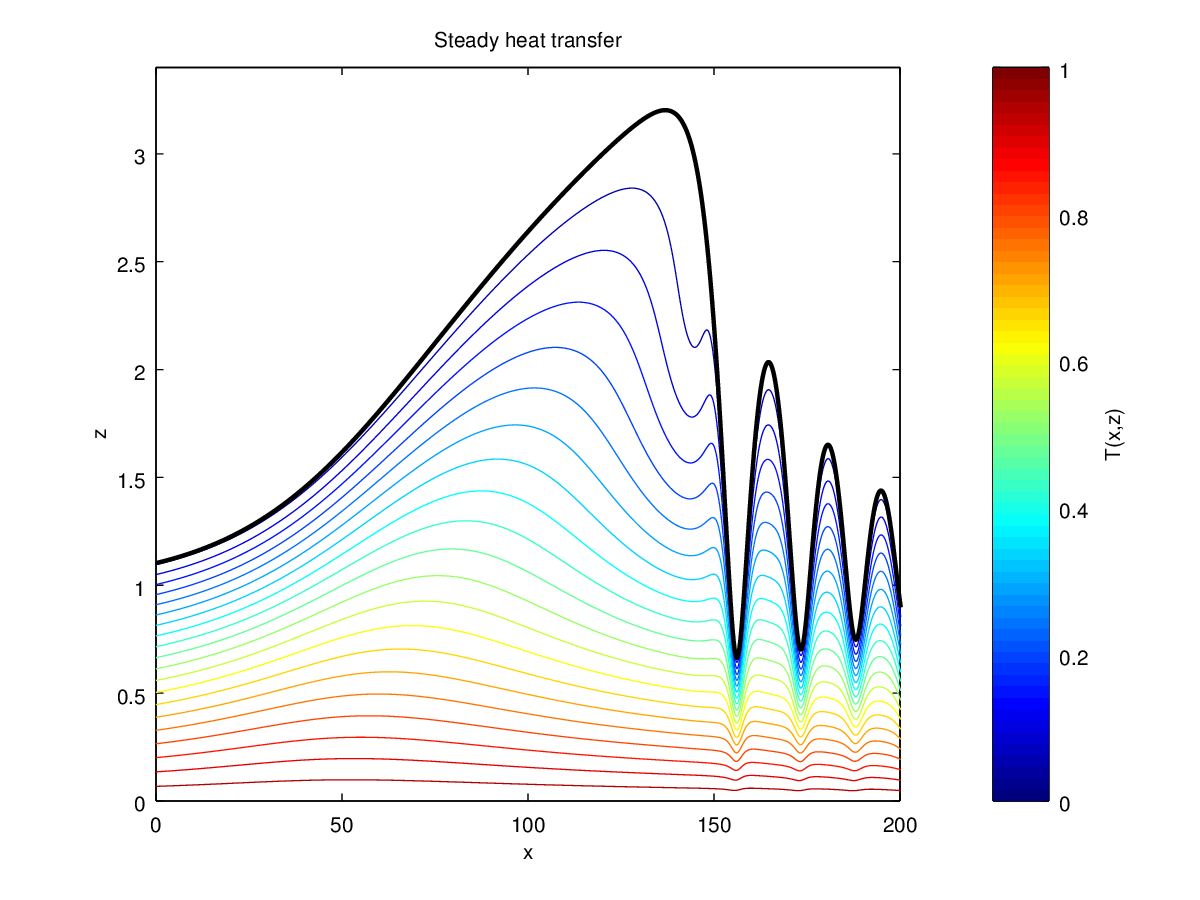}} \hfill
  \subfigure[$\PR\, =\, 30,\ \BITILDE\, =\, 0.01$]{\includegraphics[width=0.48\textwidth]{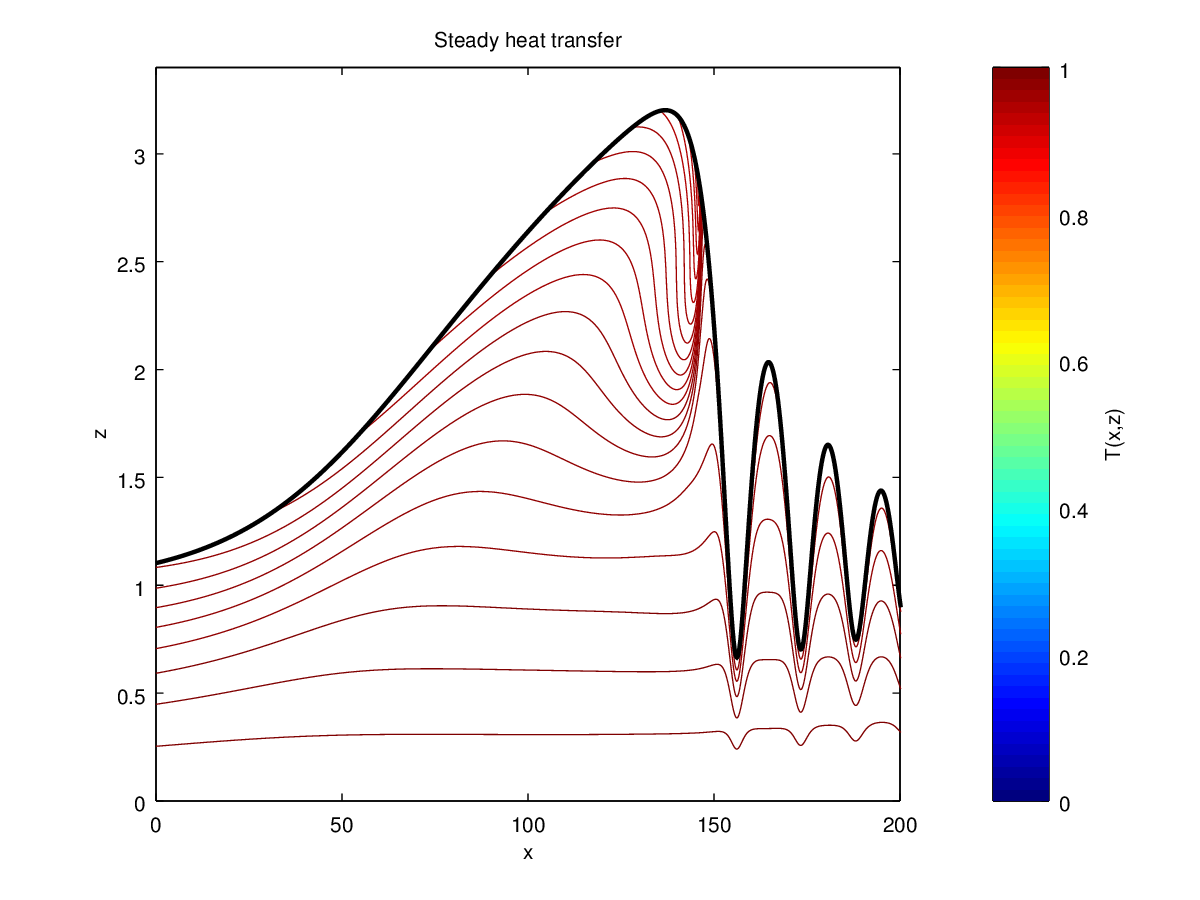}}  \hfill
  \caption{\small\em \emph{On the left column:} Isotherms in the inertial frame for different $\BITILDE$ numbers and $\PR\ =\ 7$. As expected, the temperature field is more uniform when no heat transfer is allowed with the free surface. \emph{On the right column:} Isotherms in the inertial frame for different $\PR$ number. The recirculation zone occurs as the advective part becomes dominant against the diffusive process. Heat flux becomes locally strongly non linear. $\BITILDE\ =\ 0.01\,$.}
  \label{fig:contBi}
\end{figure}

The numerical results for the steady temperature profile are obtained using an implicit second order scheme. The isothermal lines plotted in Figure~\ref{fig:contBi} have been computed from equation \eqref{fourier} for an analogous configuration as is \cite{Trevelyan2007}. The solitary wave profile and the velocity field under the wave were obtained from \textsc{Vila}'s model \cite{Boutounet2008}. The \textsc{Reynolds} number is fixed at $\RE\ =\ 7.5\,$, $\KA\ =\ 3000$ for various fluid properties ($\PR$ and $\BI\ =\ (2 \RE)^{1/3}\,\BITILDE$).

When no heat transfer is allowed between the liquid film and the surrounding air, the falling film reaches the uniform temperature given by the heated wall ($\BI\ \to\ 0$). When heat exchange between the two fluids phases is maximal ($\BI\ \to\ \infty$), the liquid film behaves as a conductive medium between the heated wall and the colder air.

The temperature follows a linear distribution across a flat film (\textsc{Nusselt} solution). However, as soon as hydrodynamic instabilities occur, a recirculation zone within large-amplitude waves may appear when the fields are described from the wave moving frame. The heat transfer through the li\-quid film is locally far from being linear. This corresponds to the physical mechanism of heat enhancement, as used in engineering process. In case of vertical falling film, \textsc{Benjamin} \cite{Benjamin1957} has shown the appearance of such inertial hydrodynamic instabilities. Therefore, when considering the vertical con\-fi\-gu\-ra\-tion of anisotherm falling film, the intensification of heat transfer by the hydrodynamic instabilities must be taken into account. The hypothesis of flat falling film does not stand anymore.

%%% ------------------------------------------------------------------------ %%%

\section{Asymptotic model}
\label{asymptotic_model}

In  what follows, we present the derivation of a system of averaged equations for the mass, momentum and heat balances \eqref{equas1} -- \eqref{equas4} following the classical \textsc{Saint}-\textsc{Venant} approach. We consider a formal expansion for the velocity, the pressure and the temperature field with respect to the order parameter $\eps\,$:
\begin{align}\nonumber
  u(x,\,z,\,t)\ &=\ u\0(x,\,z,\,t)\ +\ \eps \,u\1(x,\,z,\,t)\ +\ \O(\eps)\,, \\
  w(x,\,z,\,t)\ &=\ w\0(x,\,z,\,t)\ +\ \O(\eps)\,,\nonumber \\
  p(x,\,z,\,t)\ &=\ p\0(x,\,z,\,t)\ +\ \O(\eps)\,,\nonumber \\
  T(x,\,z,\,t)\ &=\ T\0(x,\,z,\,t)\ +\ \eps \,T\1(x,\,z,\,t)\ +\ \O(\eps^{2})\,.\label{T-expand}
\end{align}
Consistency with the long-wave asymptotic expansion of the flow variables with respect to the parameter $\eps$ is guaranteed by computing the higher order corrections to the thermal \textsc{Nusselt} solution, corresponding to the leading order term of the asymptotic development for the temperature field.

%%% ------------------------------------------------------------------------ %%%

\subsection{Thermal Nusselt solution}

The substitution of the formal development \eqref{T-expand} into the dimensionless \textsc{Fourier} equation \eqref{equas4} yields
\begin{equation*}
  \partial^2_{zz}\,T\0\ =\ 0
\end{equation*} 
by taking the limit $\eps \to 0\,$. In the following, we set
\begin{equation}\label{hyp}
  \B\ \eqdef\ \sqrt{1\ +\ (\eps\partial_x h )^2}\;\BI\ =\ \O(\BI)\,.
\end{equation}
Taking into account the associated boundary conditions
\begin{equation*}
  \left.T\0\,\right|_{z\, =\, 0}\ =\ 1\,, \qquad 
  \left.\partial_z T\0\,\right|_{z\, =\, h}\ =\ - \B\,\left.T\0\,\right|_{z\,=\, h}\,,
\end{equation*}
the resolution of the second order differential equation gives the expected expression for the main order temperature profile called the thermal \textsc{Nusselt} solution \cite{Burelbach1988}:
\begin{equation}\label{T-Nusselt}
  T\0(x,\,z,\,t)\ =\ 1\ -\ A(x,\,t)\,z
\end{equation}
with
\begin{equation}\label{equaA}
  A(x,\,t)\ =\ \dfrac{\B}{1\ +\ h(x,t)\,\B}\,.
\end{equation}
We can remark that the \textsc{Nusselt} thermal gradient $A$ is function of $h$ and its first derivative with respect to $x\,$.

Although its computation is straightforward, the expression of the thermal  \textsc{Nusselt} solution differs from the linear temperature profile found in the literature \cite{DAlessio2010, Ruyer-Quil2008}. The thermal gradient $A$ involves the liquid film height $h$ depending on the $x\,$. The proposed thermal \textsc{Nusselt} solution gives with more accuracy the influence of the hydrodynamic instabilities than as if its linear factor would just have been constant along the downward direction. It is worth to point out that, even when $\eps\ \to\ 0\,$, travelling waves may appear as solutions of the Shallow Water model. Thus, even in this limit case ($\eps\ =\ 0$), the falling film profile may vary along the downward direction and it is expected for the thermal \textsc{Nusselt} solution to describe this behaviour as well.

Remark that the average \textsc{Nusselt} solution $\bT\0\ =\ \dfrac{1}{h}\;\displaystyle \int_0^h T\0\;\ud z$ is consistent with the limit cases of heat transfer. Indeed out of the instability neighborhood, the uniform temperature profile is reached (adiabatic case: $\BI\ \to\ 0\,$, $\bT\0\ \to\ 1$) and whereas no resistance to the transfer at the free surface happens, $\BI\ \to\ \infty\,$, the linear heat profile is realized yielding $\bT\0\ \to\ \dfrac{1}{2}\,$.

%%% ------------------------------------------------------------------------ %%%

\subsection{Formal derivation of the velocity}

\begin{figure}
  \centering
  \subfigure[]{\includegraphics[width=0.49\textwidth]{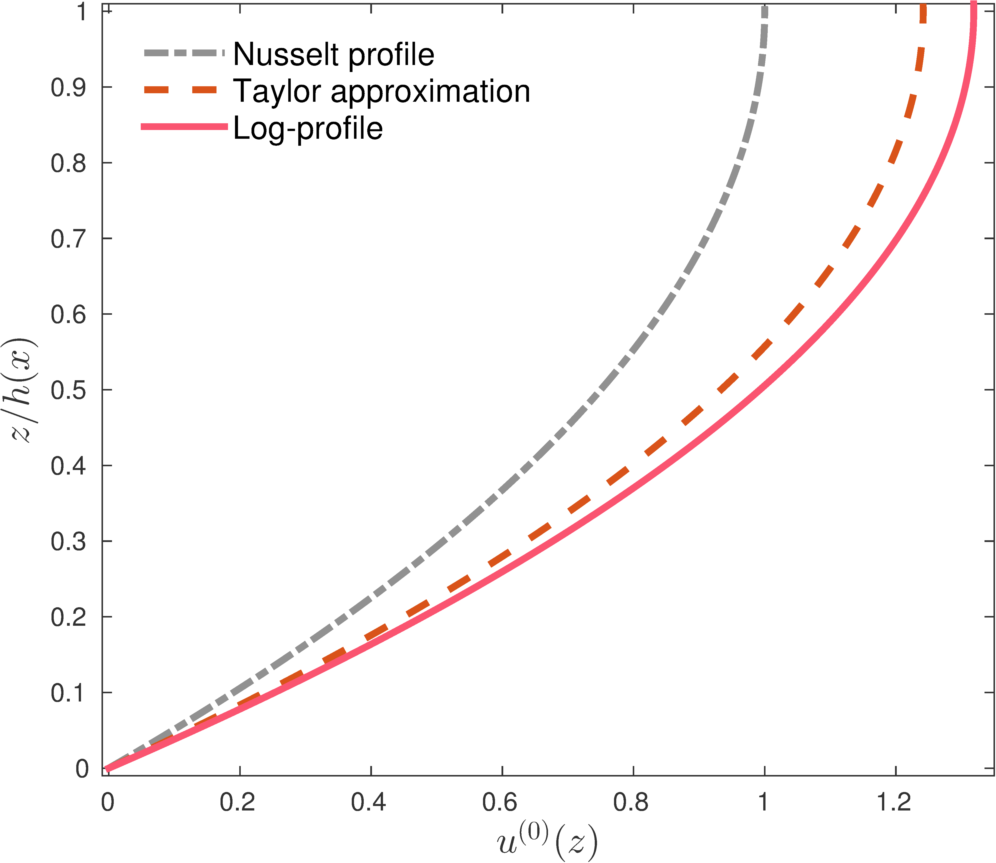}}
  \subfigure[]{\includegraphics[width=0.49\textwidth]{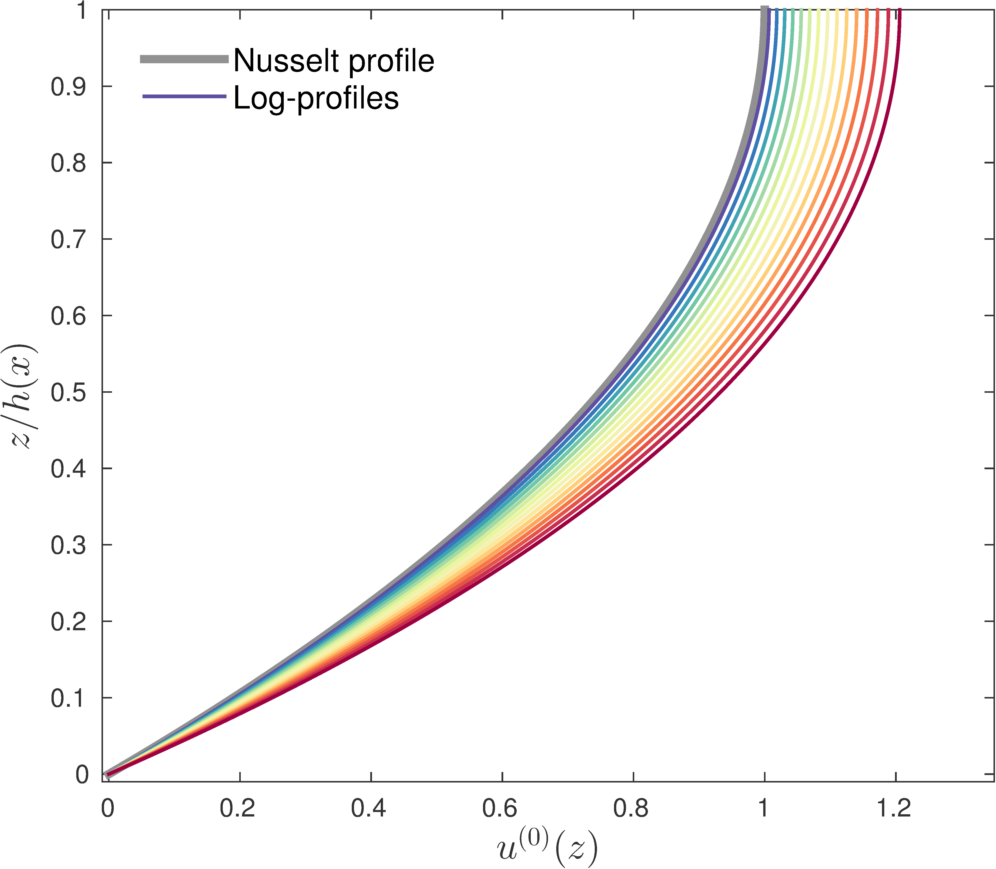}}
  \caption{\small\em Comparison of velocity profiles with $\Pi_{\mu}\ \neq\ 0\,$. Physical parameters $h\ =\ 1\,$, $A\ =\ 0.1\,$, $\Pi_{\mu}\ =\ 0.25\,$. On the right panel (b) the parameter $\Pi_{\mu}$ changes uniformly from $0.01$ to $0.25\,$.}
  \label{fig:pimu}
\end{figure}

We consider the flat-film velocity distribution, \ie $\eps\ =\ 0\,$. Integration of \eqref{equas1} using the boundary conditions \eqref{mur}, \eqref{hauteur} and \eqref{Marangoni} gives
\begin{eqnarray*}
  u^{(0)} & = & \dfrac{2}{\Pi_{\mu}A}\; \left(h\ +\ \dfrac{1\ -\ \Pi_{\mu}}{\Pi_{\mu}\,A}\right)\bigl(\ln|1 - \Pi_{\mu} + \Pi_{\mu} A z|\ -\ \ln|1-\Pi_{\mu}|\bigr)\ -\ \dfrac{2}{\Pi_{\mu}\,A}\;z  \\ & & \mbox{ if } \Pi_{\mu}\ \neq\ 0 \\
  u^{(0)} & = & 2\, h\, z - z^2\,, \qquad \Bigl(\mbox{ if } \Pi_{\mu}\ =\ 0\Bigr) \\
  u^{(0)} & = & 2\, h\, z\ -\ z^2\ +\ \left(2\, h\, z - z^2 - h\, A\, z^2 + \dfrac{2}{3}\; A\, z^3\right)\, \Pi_{\mu}  + \O(\Pi_{\mu}^2) \mbox{ if } \Pi_{\mu}\ \ll\ 1\,.
\end{eqnarray*}
The graphical comparison of different profiles is shown in Figure~\ref{fig:pimu} for an already large value of the rate of change of viscosity $\Pi_\mu\ =\ 0.25$ and a thermal gradient $A\ =\ 0.1\,$. The velocity of the flow increases with $\Pi_\mu$ as expected from the lowering of the viscosity and the subsequent reduction of the viscous stresses. A significant departure of the velocity 
profile from the \textsc{Nusselt} parabolic distribution ($\Pi_\mu\ =\ 0$) can be observed. However, the polynomial velocity distribution obtained from a \textsc{Taylor} expansion is still very close to the exact distribution which involves a logarithmic correction. This suggests that the assumption $\Pi_{\mu}\ \ \ll\ 1$ holds up to already large values of $\Pi_{\mu}\,$. Since the logarithmic correction to the velocity profile will significantly complicate the algebra and the resulting models, we will hereinafter assume only small values of the rate of change of viscosity  $\Pi_{\mu}\ \ll\ 1$ in our derivation process with the expectation that the resulting simplified equations will remain valid up to order one values of $\Pi_{\mu}$.

We thus assume $\Pi_{\mu}\ =\ \O(\eps)\,$ and introduce a constant $\varpi\ =\ \O(1)$ such that 
\begin{subequations}\label{u0-w0}
\begin{equation}\label{u0}
u^0(x,\,z,\,t)\ =\ 2h(x,\,t)\,z\ -\ z^2\ +\ \O(\eps)\,.
\end{equation}
Using the free divergence condition and the no slip condition $w(x,\,0)\ =\ 0\,$, the transverse component velocity $w$ is determined as
\begin{eqnarray}
\nonumber
  w(x,\,z,\,t)\ & = &\ -\int_0^z \partial_x u\,(x,\,y,\,t) \;\ud y \\
\nonumber 
 & = & -z^2\, \partial_x h(x,\,t)\ +\ \O(\eps) \\
  & = & w\0(x,\,z,\,t)\ +\ \O(\eps)\,.
\label{w0}
\end{eqnarray}
\end{subequations}

%%% ------------------------------------------------------------------------ %%%

\subsection{Formal derivation of the pressure}

Truncated at $\O(\eps)$, integration of the cross-tream momentum balance \eqref{equas2} yields
\begin{equation*}
  2 \,  \left.p\right|_{h}\  -2 \, p(z)= \eps \, \left.  \partial_z w\right|_{h}\ -\ \eps  \, \partial_z w(z)+  \ \O(\eps^2)\,.
\end{equation*}
The boundary condition \eqref{conditionp} gives
\begin{equation*}
  \left.p\,\right|_{h}\ =\ -\ \eps^2\,\left(\WE\ -\ \dfrac{\MA}{2}  \left.T\,\right|_{h}\right) \partial_{xx}^2 h - \eps \left.\partial_x u\,\right|_{h}\ +\ \O(\eps^2)\,.
\end{equation*}
Where the dependence of the pressure with respect to the surface tension has been made explicit even though it formally appears as a $\O(\eps^2)$ correction. As underlined above, inclusion of surface tension effects is required to capture the onset of capillary waves at the front of the solitary waves and to prevent wave breaking, which justifies the assumption of large \textsc{Weber} numbers $\WE\ =\ \dfrac{\kappa}{\eps^2}\,$. As a consequence, the 
pressure distribution at leading order reduces to the sole contribution of surface tension 
\begin{equation}\label{pression}
  p(z)\ =\ -\kappa\,\partial_{xx}^2\,h\ +\ \O(\eps)\,.
\end{equation}

%%% ------------------------------------------------------------------------ %%%

\subsection{Consistent averaged momentum equation}

In this Section, we present the derivation of a consistent averaged momentum equation following \cite{Boutounet2013}. The proposed momentum balance includes the coupling with heat transfer resulting from the dependence of surface tension and viscosity with respect to the temperature.

We first integrate the stream-wise momemtum balance \eqref{equas1} using the boundary conditions \eqref{mur} and \eqref{hauteur}, to obtain an averaged momentum balance
\begin{multline}
  \RE\;\partial_t \left(\int_0^h u\;\ud z \right)\ +\ \RE\,\partial_x \left(\int_0^h u^2\;\ud z \right)\ =\ \dfrac{2h}{\eps}\ -\ 2\,\partial_x \left( \int_0^h p\;\ud z \right) \\ 2\,\left.p\,\right|_{h}\,\partial_x h\ +\ \dfrac{1\ -\ \eps\,\varpi\,T}{\eps}\left.\partial_z u\,\right|_{h}\ -\ \dfrac{1\ -\ \eps\,\varpi\,T}{\eps}\left.\partial_z u\,\right|_{0}\ +\ \O(\eps)\,.\label{av-mom}
\end{multline}
We know that $u\ =\ u^0\ +\ \eps u^1\ +\ \O(\eps^2)$ with $u^0(x,z,t)\ =\ 2\,h(x,\,t)\,z\ -\ z^2$ and
\begin{multline*}
  u^1\ =\ -\MA\,\dfrac{\ud T^0(h)}{\ud x}\,z\ +\ \varpi \left[\,2\,z\,h\ -\ z^2\,(1\ +\ A)\,h\ +\ \dfrac{2}{3}\;z^3\,A\,\right]\\ +\ \dfrac{z\,h}{6}\;\RE\,\left[\,8\, h^3\ -\ 4\, z^2\, h\ +\ z^3\right]\, \partial_x h
\end{multline*}
where $\dfrac{\ud T^0(h)}{\ud x}\ =\ -(h\,\partial_h A\ +\ A)\, \partial_x h\,$.

We introduce the depth-averaged velocity
\begin{equation}
  v(x,\,t)\ =\ \dfrac{1}{h} \displaystyle\int_0^h u(x,z,t) \,\ud z\ =\ \dfrac{2}{3}h^2(x,t)\ +\ \O(\eps)\label{v}
\end{equation}
so that 
\begin{equation}\label{u2}
  \displaystyle \int_0^h u^2\;\ud z\ =\ \dfrac{8}{15}\;h^5\ +\ \O(\eps)\ =\ v^2\, h\ +\ \dfrac{4}{45}\,h^5\ +\ \O(\eps)\,.
\end{equation}
Equation \eqref{hauteur} gives
\begin{equation}\label{hauteurnew}
  \partial_t h=- \partial_x\,(h v) = -2h^2 \partial_x h + \O(\eps) \,.
\end{equation}
The boundary condition \eqref{Marangoni} gives
\begin{equation}
  \dfrac{1 - \eps  \, \varpi\,T}{\eps}\left. \partial_z u\,\right|_{z\,=\,h}\  =\ -\MA\;\dfrac{\ud T^0(h)}{\ud x}\ +\ \O(\eps)\label{uzh}
\end{equation}
whereas the wall shear stress is evaluated from the asymptotic expansion which leads to 
\begin{multline}
  \dfrac{1 - \eps\,\varpi\,T}{\eps}\left. \partial_z u\, \right|_{z\,=\,0}\ =\ \dfrac{3}{\eps}\;\left[\,1\ -\ \eps\,\varpi\,\dfrac{3\ +\ T^0(h)}{4}\,\right]\dfrac{v}{h}\\
  -\dfrac{4}{15}\;\RE\; h^4\; \partial_x h\ +\ \dfrac{1}{2}\;\dfrac{\ud T^0(h)}{\ud x}\ -\ 2\,\kappa\,h\,\partial_{xxx}^3\,h\ +\ \O(\eps)\,.\label{uz0}
\end{multline}

The chosen expressions \eqref{u2} and \eqref{uz0} of the momentum flux and of the wall shear stress are not unique as one can play with the asymptotic expression of the averaged speed \eqref{v}. Equation \eqref{u2} introduces the classical momentum flux $v^2\,h$ of the shallow-water equations corrected with a term $4\,h^5\,/\,45$ that is function only of the film thickness. Following \cite{Boutounet2013}, where the analogy of the shallow-water equations with the compressible \textsc{Euler} equations is underlined, the film thickness $h$ being the analogue of the density, the correction $4\,h^5\,/\,45$ is a barotropic contribution to the pressure. The wall shear stress expression \eqref{uz0} is chosen so that the dependency of the viscosity with respect to the temperature appears as a correction to $3\,v\,/\,h$ which corresponds to the classical \textsc{Nusselt} parabolic profile.

The asymptotic expression of the pressure distribution \eqref{pression} gives
\begin{equation}\label{p}
  -2\,\partial_x \left(\int_0^h p\;\ud z \right)\ +\ 2\,\left. p\,\right|_{h}\, \partial_x h\ =\ 2\,\kappa\,h\,\partial_{xxx}^3\,h\ +\ \O(\eps)\,.
\end{equation}
From \eqref{u2}, \eqref{uzh}, \eqref{uz0} and \eqref{p}, the averaged momentum balance \eqref{av-mom} reads
\begin{multline}
  \RE\,\left(\,\partial_t (h \, v)\ +\ \partial_x \left(h \, v^2\ +\ \dfrac{8}{225}\;h^5\,\right)\,\right)\ = \\ \dfrac{1}{\eps}\;\left\{2h\ -\ 3\,\left[\,1\ -\ \eps\,\varpi\;\dfrac{3\ +\ T^0(h)}{4}\,\right]\;\dfrac{v}{h} \right\}\\
  -\ \dfrac{3}{2}\;\MA\,\dfrac{\ud T^0(h)}{\ud x}\ +\ 2\,\kappa\,h\,\partial^3_{xxx}\,h\ +\ \O(\eps)\label{ouf} 
\end{multline}

The asymptotic momentum balance \eqref{ouf} extends the result obtained by \cite{Boutounet2013} by including the principal coupling terms with the heat transfer, \ie a reduced wall friction and a driving stress proportional to the gradient of free surface temperature, which accounts for the \textsc{Marangoni} effect.

%%% ------------------------------------------------------------------------ %%%

\subsection{Asymptotic heat transfer model}

Before turning to the averaging of the heat equation \eqref{equas4}, we first compute the second-order correction  $T\1$ to the linear distribution $T\0$ given by \eqref{T-Nusselt}. The substitution of $T\ =\ T\0\ +\ \eps\, T\1\,$, $u\ =\ u\0\ +\ \O(\eps)$ and $w\ =\ w\0\ +\ \O(\eps)$ gives
\begin{equation*}
  \partial_{zz} T\1\ =\ \PE \left(\partial_t T\0\ +\ u\0 \partial_x T\0\ +\ w\0 \partial_z T\0 \right)\,,
\end{equation*}
which can be easily integrated with the help of the boundary conditions
\begin{equation*}
  \left.\partial_z T\1\,\right|_{h}\ =\ - \B \left.T\1\,\right|_{h} \qquad \left.T\1\right|_{0}\ =\ 0\,,
\end{equation*}
and the expressions \eqref{u0-w0} of $u\0$ and $w\0\,$. We thus obtain
\begin{eqnarray*}
 T\1 (x,\,z,\,t)\ &=& \frac{\PE\,\B\,z\partial_x h}{ 60(1\ +\ h\,\B)^3}\;\left\{ z^3 [5 - z\,\B] + h \left[ z^3 \B (20 - 3 z \B) \right.\right. \\&&  
  + h \left(5z^2\B \left(-4+3z\B)+ 2h (-10(1 +z^2\B^2)  \right.\right. \\ 
  &&  \left. \left.  \left. \left.+ h\,\B (5 +4h\,\B)
  \right)\right) \right] \right\}
\end{eqnarray*}

%%% ------------------------------------------------------------------------ %%%

\subsection{The conservative formulation of the model}

In this section, we derive formally an evolution equation for the free-surface temperature temperature following the consistent \textsc{Saint}-\textsc{Venant} approach.

We introduce the free surface temperature
\begin{equation}\label{def-bt}
   \bt\ \eqdef\ =\ T(z=h)
\end{equation}
so that the temperature field may be written
\begin{equation*}
  T\ =\ 1 -(1-\bt)\,\frac{z}{h}\ +\ \eps{\tilde T\1}\ =\ \tilde T\0\ +\ \eps{\tilde T\1}\,,
\end{equation*}
where the $\O(\eps)$ correction $\tilde T\1$ verifies $\tilde T\1(z=h)\ =\ 0$ as required by the definition \eqref{def-bt} so that $\tilde T\1\ =\ T\1\ -\ T\1(z=h)\ +\ \O(\eps^2)\,$.

Integration of the the heat equation \eqref{equas4} across the film height yields
\begin{equation}\label{eq-int}
  \partial_t \int_0^h T \,\ud z\ +\ \partial_x\int_0^h u\,T \,\ud z  =- \dfrac{1}{\eps\,\PE} \left(   \B\,\left.T\,\right|_{z\, =\, h} + \left.\partial_z T\,\right|_{z\, =\, 0} \right) +\ \O(\eps) \,,
\end{equation}
which involves the average temperature $h^{-1}\int_0^h T \,\ud z$  and the mixing temperature $q^{-1}\int_0^h u\,T \,\ud z$. The average temperature $h^{-1}\int_0^h T \,\ud z$ can be easily derived from the free-surface temperature $\bt$ through the linear temperature distribution \eqref{T-Nusselt} by $h^{-1}\int_0^h T \,\ud z = (1 +\bt)/2 + \O(\eps)$. An expression of the mixing temperature $q^{-1}\int_0^h u\,T \,\ud z$ in terms of the variables $h$, $q$ and $\bt$ requires a closure which can be provided again from the \textsc{Nusselt} solution  $q^{-1}\int_0^h u\,T \,\ud z = q^{-1}\int_0^h u\0 \,\tilde T\0 \,\ud z + \O(\eps) = \frac{1}{8} \left(3 + 5 \bt\right) + \O(\eps)$. Yet, this closure is not unique as one can play with the asymptotic expressions of the free-surface temperature and flow rate 
\begin{subequations}\label{q0-bt0-bt1}
\begin{eqnarray}
  q&=& q\0 +  \O(\eps) = \frac{2}{3} h^3 + \O(\eps)\,,\\
  \bt &=& \bt\0 + \eps \bt\1 +  \O(\eps^2) \nonumber \\
  &=& \frac{1}{1 + h\,\B} + \eps\frac{\PE\,\B\,h^4 \partial_x h (-15 + 7h\,\B)}{60(1+ h\,\B)^3} + \O(\eps^2)\,.
\end{eqnarray}
\end{subequations}
The point of view that is adopted below is to express the convective terms at the l.h.s. of \eqref{eq-int} as a function of $h\left(\partial_t \bt\ +\ v \partial_x \bt\right)\ =\ \partial_t(h\bt)\ +\ \partial_x(q \bt)\,$. This choice stems from i) the classical form of the shallow water equation and the choice of $h\,$, $q\ =\ hv$ anf $\bt$ as the variables of the model, ii) the fact that the hyperbolic structure of the obtained evolution equations is then guaranteed as will be underlined in Section~\ref{sec:num}. The r.h.s. of \eqref{eq-int} is written as a function of $\bt$ using
\begin{eqnarray*}
  \B\,\left.T\,\right|_{z\, =\, h} + \left.\partial_z T\,\right|_{z\, =\, 0} &=&  \B\,\left.\tilde T\0 \,\right|_{z\, =\, h}\ +\ \left.\partial_z \tilde T\0\,\right|_{z\, =\, 0} + \O(\eps)\\
  &=& \frac{1 +h\,\B}{h}\;\left(\bt - \bt\0\right)\ +\ \O(\eps)
\end{eqnarray*}
We thus write the following balance 
\begin{equation}\label{eq-bt-1}
  \frac{1 +h\,\B}{h}\;\left(\bt - \bt\0\right)\ =\ - \left(\frac{3}{16} -\frac{7}{80} h\B\right) \eps\,\PE \left\{\partial_t(h\bt)\ +\ \partial_x(q \bt) \right\}
\end{equation}
where the coefficients $a\ =\ 3/16$ and $b\ =\ -7/80$ are adjusted to verify consistency at $\O(\eps)$ after substitution of the expansions \eqref{q0-bt0-bt1}. An alternative to \eqref{q0-bt0-bt1} is offered by $\bt = \bt\0 + \O (\eps)$ as $\frac{3}{16}\ -\ \frac{7}{80}\;h\B\ =\ \frac{11}{40}\ -\ \frac{7}{80}/\bt\0$ so that
\begin{equation}\label{eq-bt-2}
  \frac{1 + h\,\B}{h}\;\left(\bt - \bt\0\right)\ =\ - \eps\,\PE\left\{\partial_t[h\,g(\bt)]\ +\ \partial_x[q\,g(\bt)] \right\}
\end{equation}
with
\begin{equation}\label{eq-bt-3}
  g(\bt)\ =\ \frac{11}{40}\;\bt\ -\ \frac{7}{80}\;\ln(\bt)
\end{equation}
is also consistent at $\O(\eps)\,$.

Before turning to the complete formulation of our coupled set of transport equations for the variables $h\,$, $q\ =\ hv$ and $\theta\,$, let us comment the results \eqref{eq-bt-1} and \eqref{eq-bt-2}. First the convection terms at the r.h.s. of \eqref{eq-bt-1} cancel out at a specific value of the film height $h_c\ \eqdef\ 15/(7\B)\,$. Similarly, the r.h.s. of \eqref{eq-bt-2} vanish at a given value of the free-surface temperature $\bt_c\ \eqdef\ 7/22\,$. It is easy to show that these two critical values are related by $\bt_c\ =\ \bt\0(h_c)\,$. Worse, for $h$ above $h_c$ or $\bt$ below $\bt_c$, the relaxation term at the l.h.s. of \eqref{eq-bt-1} and \eqref{eq-bt-2} becomes an amplification one. Obviously, these critical values are nonphysical and thus present clear limitations to the applicability of the averaged heat equations \eqref{eq-bt-1} and \eqref{eq-bt-2} to only relatively low values of $\B\,$.

A solution to this problem may be found by playing with the equivalence $\bt\ =\ \bt\0\ +\ \O(\eps)\,$. We may rewrite the function $g$ appearing in \eqref{eq-bt-3} as
\begin{eqnarray}
\nonumber
 \tilde g(\bt,h)\ &=&\ \frac{11}{20}\left[\bt- 2\ln(\bt) -\frac{1}{2}\bt\0 + \frac{81}{44} \ln(\bt\0)\right] \\
& =& =  g(\bt)\ +\ \O(\eps)\label{eq-bt-4}
\end{eqnarray}
This particular choice of $\tilde g$ stems from the requirements that (i) $\tilde g$ remains differentiable for all acceptable values of the variables $h$ and $\theta\,$, annd that (ii) $\theta$ remains in the unit interval $[\,0,\,1\,]\,$.

Indeed, let us consider a solitary wave of infinite extension relaxing at both ends to the flat film solution $h\ =\ 1\,$. An example of such waves has been shown in Figure~\ref{fig:contBi}. In the moving frame of coordinate $\xi\ =\ x\ -\ \,c\, t\,$, $c$ referring to the phase speed of the wave, integration of the mass balance gives $q\ =\ c\,h\ +\ q_0\,$, where $q_0\ =\ \int_0^h u\,\ud z$ is the constant flow rate under the wave. Equation \eqref{eq-bt-2} with \eqref{eq-bt-4} then simplifies into
\begin{equation}\label{eq-bt-5}
  q_0\;\frac{d}{d\xi}{\tilde g}\ =\ \O(1/\PE)
\end{equation}
So that $\tilde g$ is nearly constant along the wave. As the wave departs from and returns to the \textsc{Nusselt} flat film solution $h\ =\ 1\,$, one expects $\tilde g\ \approx\ \tilde g(\bt\0(h=1),\,1)\,$. A necessary condition to insure $0\ <\ \theta\ <\ 1$ consists in verifying that, the solutions to  
\begin{equation}\label{eq-bt-6}
  \tilde g(\bt,h)\ =\ \tilde g(\bt\0(h=1),\,1)
\end{equation}
lay within this interval for all admissble values of $h$ and $\B$. An expansion of \eqref{eq-bt-6} for $\B\ \ll\ 1$ gives $\bt\ =\ 1\ -\ \frac{59}{44}\;\B\,\left(h\,-\,\frac{15}{59}\right)\ +\ \O(\B)\,$. Therefore, in this limit, the solution to \eqref{eq-bt-6} remains in the unit interval provided that $h\ >\ 15/59\ \approx\ 0.25\,$, a condition that is satisfied for all considered solitary waves. Conversely, in the limit $\B\ \gg\ 1\,$, $\bt\ \approx\ \left(\bt\0\right)^{81/88}$ so that the solution $\bt$ to \eqref{eq-bt-6} lays in the unit interval $[\,0,\,1\,]\,$.

%%% ------------------------------------------------------------------------ %%%

\subsection{Vectorial form of the model}

The system of equations \eqref{hauteurnew}, \eqref{ouf}, \eqref{eq-bt-2} with \eqref{eq-bt-4}  can be written in vectorial form
\begin{equation}\label{eq:fah}
  \partial_t\,\mathcal U\ +\ \partial_x\,\mathcal F\ =\ \dfrac{1}{\eps}\;\mathcal N\ +\ \mathcal S\ +\ \O(\eps)
\end{equation}
with 
\begin{equation*}
    U\ =\ \begin{pmatrix}
    h \\
    \\
    h\, v \\
    \\
    h\,\phi
  \end{pmatrix}\,,
 \qquad
 \mathcal F\ 
  =\ \begin{pmatrix}
   h\, v \\
   \\
   h\,v^2\ +\ \dfrac{8}{225}\;h^5\ -\ \dfrac{3\,\MA}{\RE}\;\bt \\
   \\
   h\,v\phi
  \end{pmatrix}
\end{equation*}
where $\phi\ =\ \bt\ -\ 2\ln(\bt)\ -\ \frac{1}{2}\;\frac{1}{1+\B h}\ +\ \frac{81}{44}\;\ln\left(\frac{1}{1+\B h}\right)$ and
\begin{equation*}
  \mathcal N\ =\ \begin{pmatrix}
  0 \\
  \\
  \dfrac{1}{\RE}\;\left\{2h\ - 3 \left[1 - \eps\,\varpi\, \left(1\, -\,\dfrac{\bt}{2}\right)\right]\,\dfrac{v}{h}\right\} \\
  \\
  \dfrac{1}{\PE}\dfrac{20}{11\,h}\;\Bigl(1 - (1\,+\,h\,\B)\;\bt\Bigr)
  \end{pmatrix}\,,
  \qquad
  \mathcal S\ =\ \begin{pmatrix}
  0 \\
  \\
  \dfrac{2 \, \kappa}{\RE}\; h\,\partial_{xxx}^3 h \\
  \\
  0
  \end{pmatrix}\,.
\end{equation*}

The obtained system of equations must be contrasted with the non-conservative system corresponding to the averaged energy balance derived in \cite{Trevelyan2007} following the weighted residual approach introduced by \cite{Ruyer-Quil2005}:
\begin{equation}\label{Ruy05}
  \partial_t\bt\ +\ \frac{27}{20}\, v\,\partial_x \bt\ -\ \frac{7}{40}\;\frac{1\ -\ \bt}{h}\;\partial_x\,(v h)\ =\ \dfrac{1}{\eps\,\PE}\;\frac{3}{h^2}\;\left[1\ -\ \bigl(1\ +\ \B\,h\bigr)\bt\right] 
\end{equation}
This equation can be recasted in the vectorial form of \eqref{eq:fah} with the following third components of vectors $\mathcal{S}$ and $\mathcal{N}$ given by 
\begin{equation*}
  \mathcal{N}_3\ =\ \dfrac{1}{\PE}\;\dfrac{3}{h}\;\Bigl(1 - (1\,+\,h\,\B)\;\bt\Bigr)
  \quad \hbox{and}\quad
  \mathcal{S}_3\ =\ \dfrac{7}{40}\;(1\,-\,\bt)\;\partial_x\,(hv)\ -\ \dfrac{7}{20}\;h\,v\,\partial_x\,\bt\,.
\end{equation*}
The two descriptions of the energy balance are consistent at $\O(\eps)$ and therefore shall yield the similar results as long as the long-wave expansion strictly holds, \ie for $\varepsilon \PE\ \ll\ 1\,$. However, at $\PE\ =\ \O(1)$, discrepencies shall be observed owing to the different mathematical structures of the two balances. Indeed, equation~\eqref{Ruy05} has the disadvantage not to admit a conservative form.

%%% ------------------------------------------------------------------------ %%%

\section{Numerical illustrations}
\label{sec:num}

In this Section, we study first the time-space behaviour of the asymptotic model for the average temperature field $\bT$. A hyperbolic scheme is implemented in order to take the full advantage of the conservative formulation of model \eqref{eq:fah}. A second numerical test is performed, illustrating the global behaviour of the model in function of the physical parameters $\RE$, $\BI$, $\PE\,$. The solutions representation corresponds to the dynamical system point of view, and is realized thanks to the \texttt{AUTO07p} software \cite{Doedel2012}. Notice that for the sake of simplicity in all computations below we adopt the assumption \eqref{hyp}.

%%% ------------------------------------------------------------------------ %%%

\subsection{Unsteady simulations}

The hyperbolicity of system \eqref{eq:fah} can be easily checked by computing the eigenvalues of the advective flux $\mathcal F(\mathcal U)$ \textsc{Jacobian} matrix:
\begin{eqnarray*}
  \mathcal{A}&=& \pd{\mathcal F}{\mathcal U}\\
 &=& 
  \begin{pmatrix}
    0 & & 1 & & 0 \\
    -v^2\ +\ \dfrac{8}{45}\;h^4\ +\ \dfrac{3\,\MA}{\RE\, h(1-2/\bt)}\,\left(\phi -
 \dfrac{\B h(59+81\B h)}{(1+\B h)^2}\right) & & 2\, v & &  -\dfrac{3}{h(1-2/\bt)}\;\dfrac{\MA}{\RE} \\
    -\phi\, v & & \phi & & v
  \end{pmatrix}.
\end{eqnarray*}

The eigenvalues of the matrix $\mathcal{A}$ can be computed. We obtain three distinct eigenvalues:
\begin{equation*}
  v,\quad v\ \pm\ \dfrac{2\, \sqrt{10}}{15}\;h^2 + \O(\MA/\RE)
\end{equation*}
Thus, system \eqref{eq:fah} has a very interesting property: it is always hyperbolic. The eigenstructure of the advective flux $\mathcal F(\mathcal U)$ will be used below to solve numerically the system \eqref{eq:fah} of balance laws with the widely-known \textsc{Rusanov} scheme. In order to solve numerical the conservative part of thermal \textsc{Saint}-\textsc{Venant} equations we employ the standard finite volume discretization \cite{Dutykh2010e} along with the \textsc{Rusanov} scheme \cite{Rusanov1962}. The dispersive (\ie the capillary force) and other non-conservative terms were discretized using the central finite differences. Good numerical properties of this combination were demonstrated in \cite{Noble2014}. For the time discretization we use the variable order \textsc{Adams}--\textsc{Bashforth}--\textsc{Moulton} predictor-corrector solver, which is implemented in \texttt{Matlab} in \texttt{ode113} routine \cite{Shampine1997}. The absolute and relative tolerances were both set to $10^{-6}$ in simulations shown below.

\begin{figure}
    \centering
    \includegraphics[width=0.49\textwidth]{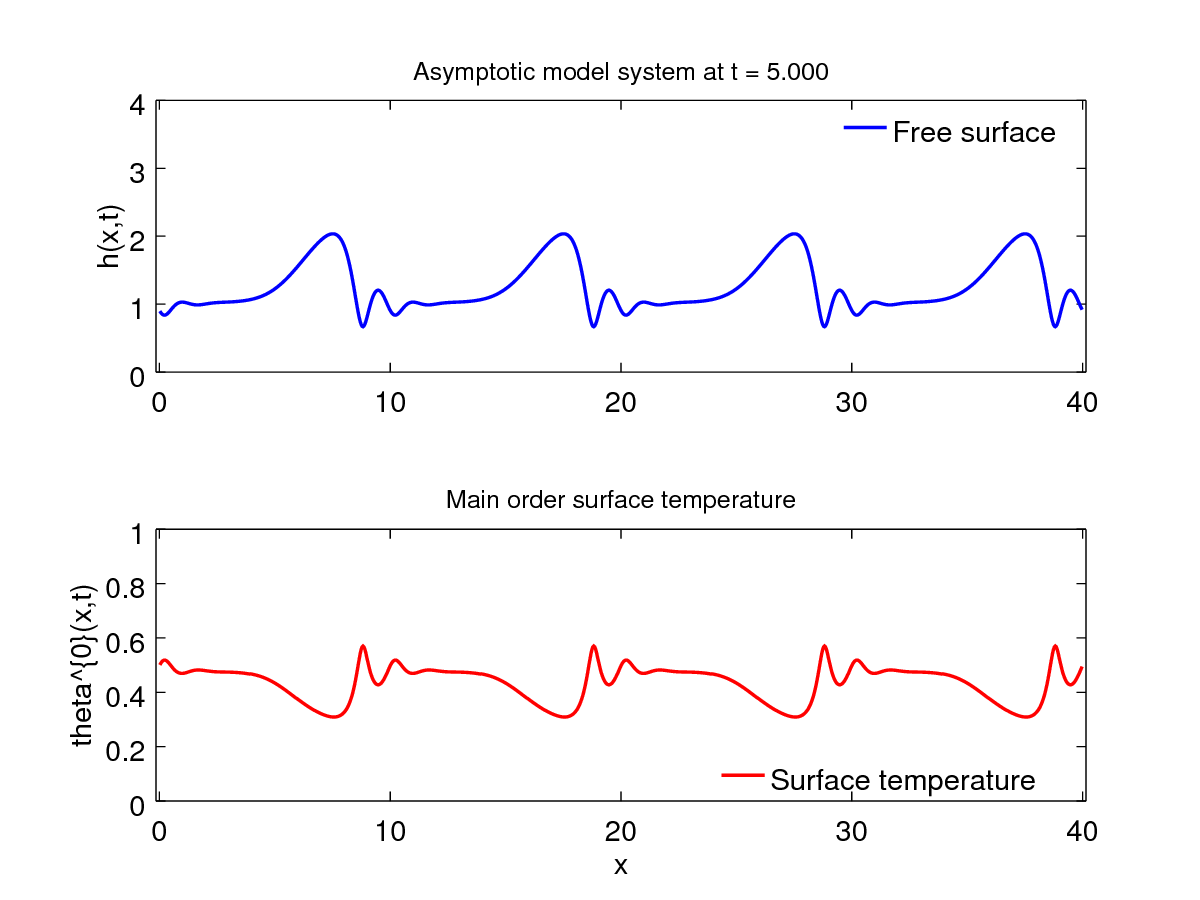}
    \includegraphics[width=0.49\textwidth]{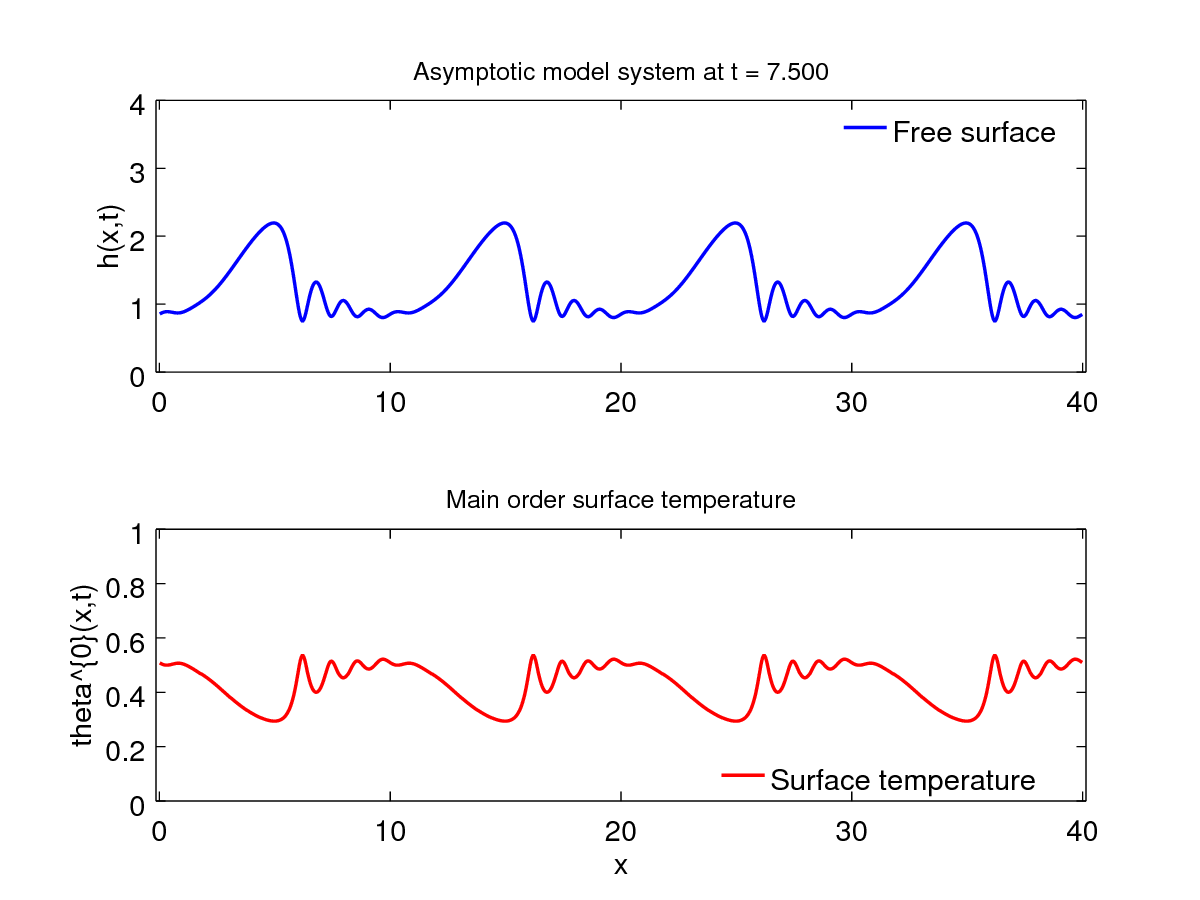}
    \includegraphics[width=0.49\textwidth]{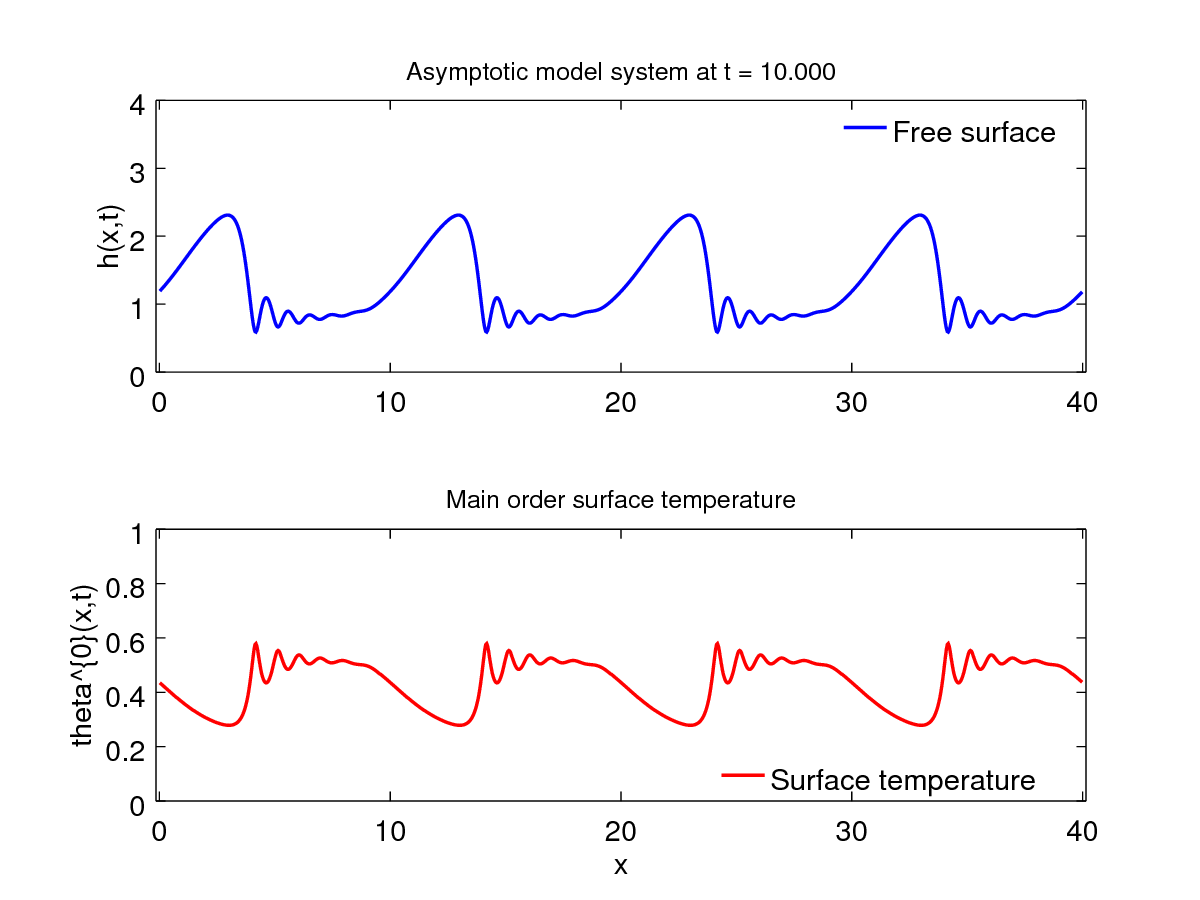}
    \includegraphics[width=0.49\textwidth]{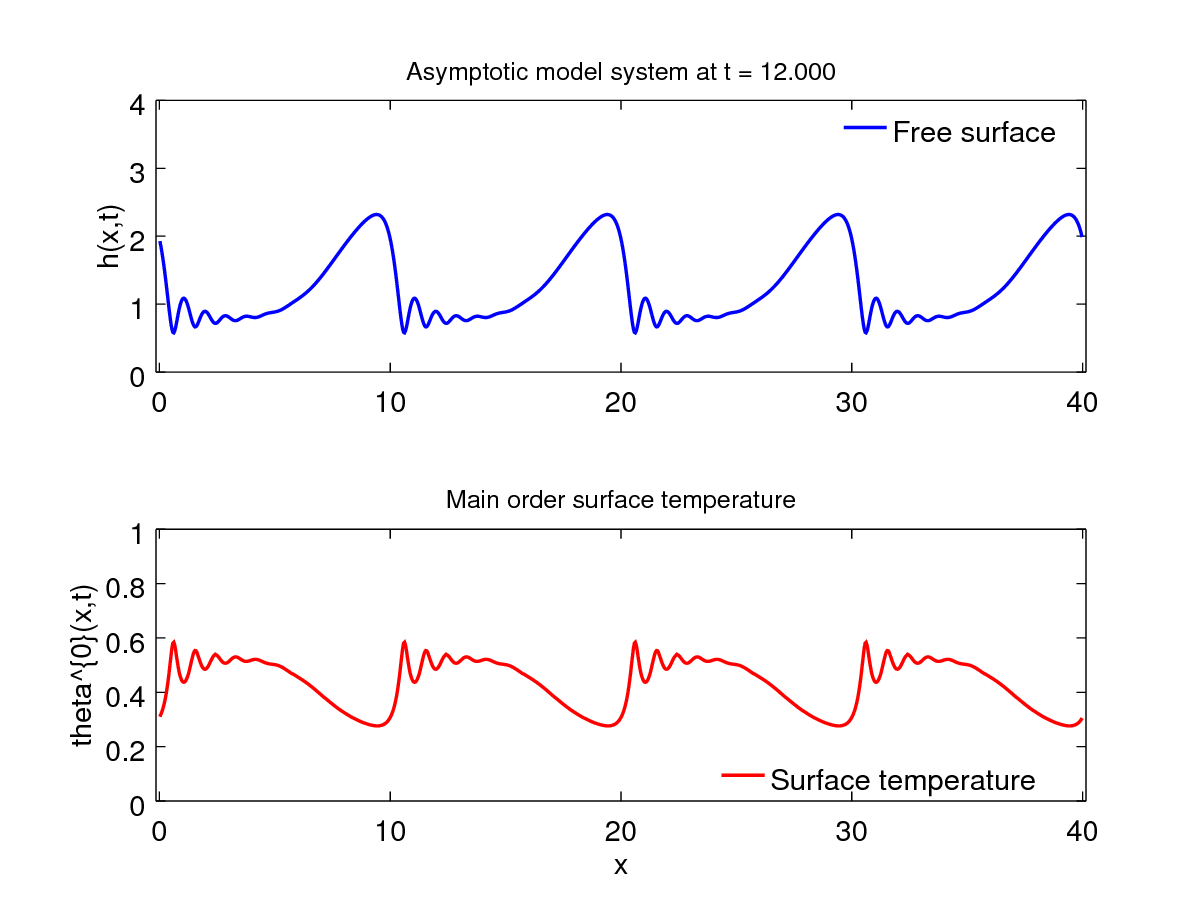}
    \includegraphics[width=0.49\textwidth]{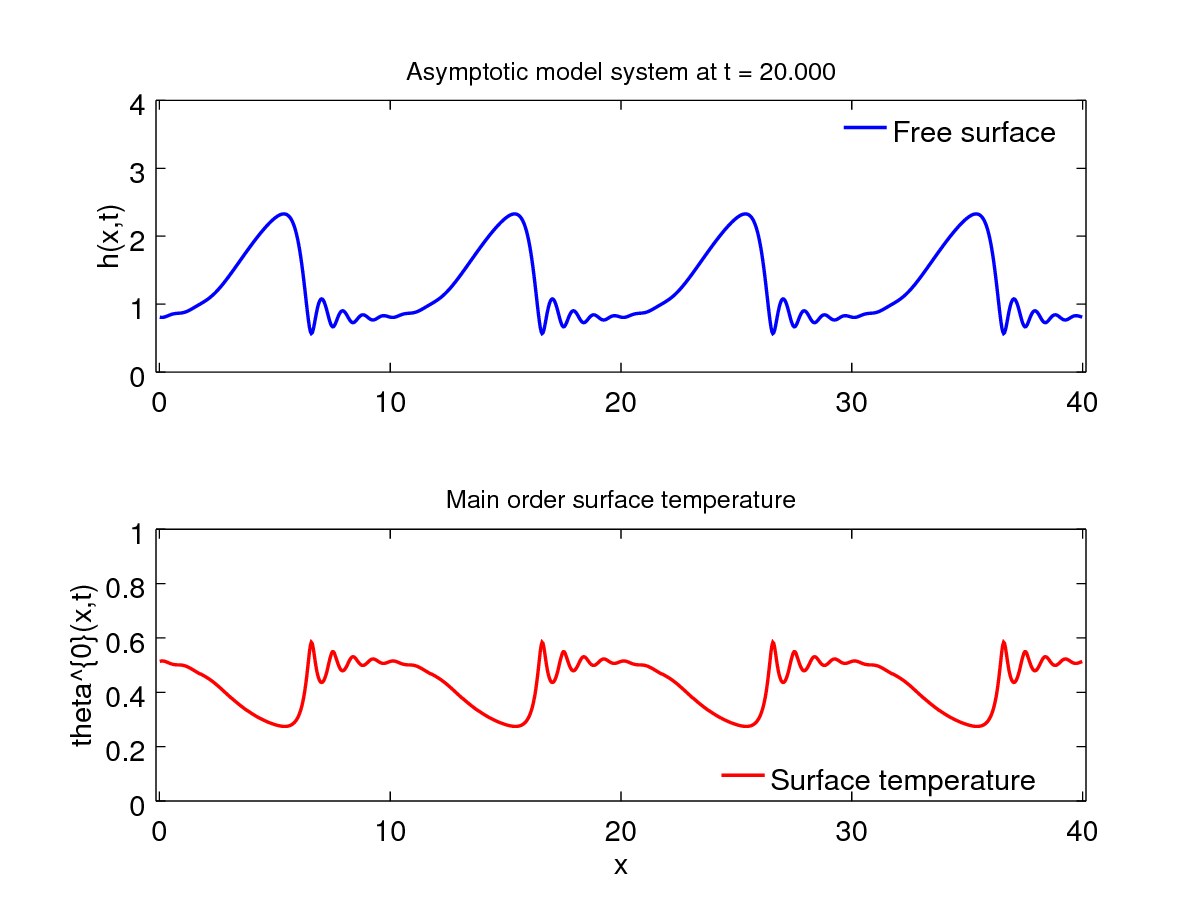}
    \includegraphics[width=0.49\textwidth]{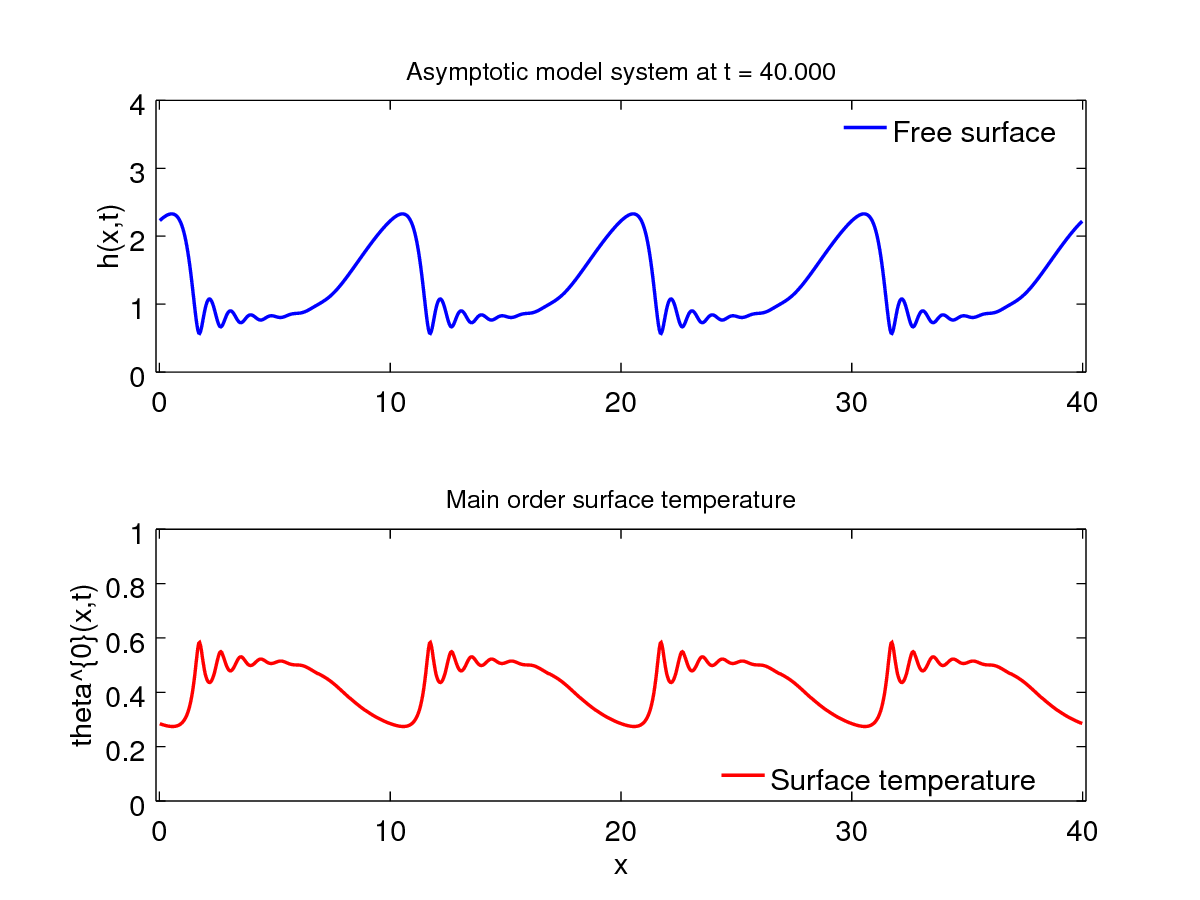}
    \caption{\small\em Time evolution of hydrodynamic free surface and first-order surface temperature. Dimensionless parameters are: $\RE\ =\ 1\,$, $\PE\ =\ 10\,$, $\BI\ =\ 1\,$, $\MA\ =\ 0\,$, $\Pi_\mu\ =\ 0\,$.}
    \label{fig:sample_subfiguresMa0Pm0}
\end{figure}

We consider the case of decoupled hydrodynamics and heat transfer ($\MA\ =\ 0\,$, $\Pi_\mu\ =\ 0\,$). The parameters are $\B\ =\ 1\,$, $\RE\ =\ 1\,$, $\PE\ =\ 10\,$ and the results of the simulation are depicted in Figure~\ref{fig:sample_subfiguresMa0Pm0}. Starting with an arbitrary reasonable initial condition
\begin{equation*}
  h_0(x)\ =\ 1\ +\ \frac{1}{2}\,\sin^2\Bigl(\frac{25\, \pi\, x\, L}{10000}\Bigr),
\end{equation*}
and $\bt^{(0)}_0\ =\ \left(1\ +\ h_0 \BITILDE\right)^{-1}$ and using periodic boundary conditions, the solution reaches its hydrodynamic steady-state at around $t\ =\ 10\,$, whereas the steady state thermal regime takes place a little bit after at around $t\ =\ 12\,$, as a result of a weaker thermal diffusivity in comparison to the kinematic viscosity ($\Pr\ =\ 10$). The mean temperature $\bT$ closely follows the free surface evolution with a minimal value reached at the crests of the waves.

%%% ------------------------------------------------------------------------ %%%

\subsection{Dynamical system representation}

Travelling-wave solutions to \eqref{eq:fah} have been looked after using the \texttt{AUTO07p} software \cite{Doedel2012}. The system of partial differential equations \eqref{eq:fah} simplifies into ordinary differential equations in the moving frame of reference, $\xi\ =\ x\ -\ c\,t\,$, where $c$ refers to the phase speed of the waves. The averaged velocity $v\ =\ c\ +\ q_0\,/\,h$ is computed after integration of the mass balance, where $q_0\ =\ \displaystyle\int_0^h (u\ -\ c)\,\ud z$ is the conserved flow rate in the moving frame. After elimination of $v\,$, one is led to a single ODE which is next recast as an autonomous dynamical system in a four-dimensional phase space spanned by $U\ =\ \bigl(h,\, \ud h/\ud\xi,\,\ud^2h/\ud \xi^2,\,\theta\bigr)\,$. Travelling waves correspond to limit cycles in the phase space which arise from Hopf bifurcations of the \textsc{Nusselt} solution $U\ =\ \bigl(1,\,0,\,0,\,\B\,/\,[2\,(1\,+\,\B)]\bigr)\,$. Solitary waves are next found through homoclinic bifurcations by increasing the period of the limit cycles. The procedure is detailed in \cite{Kalliadasis2012}.

\begin{figure}
  \centering
  \subfigure[$\PR\,=\,7$, $\BITILDE\, =\, 0.01$]{\label{fig-AutoA}
  \includegraphics[width=0.48\textwidth]{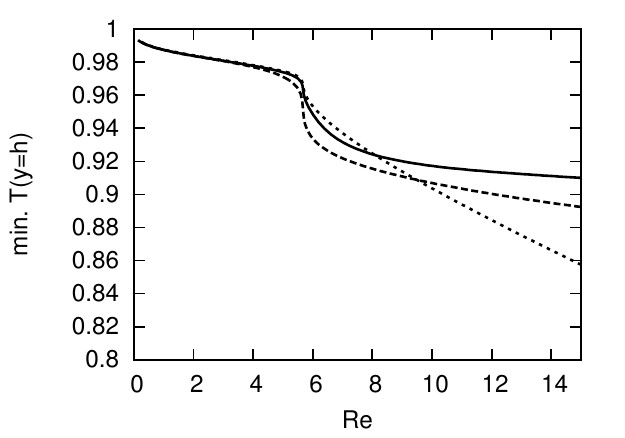}}
  \subfigure[$\PR\,=\,7$, $\BITILDE\, =\, 0.1$]{\label{fig-AutoB}
  \includegraphics[width=0.48\textwidth]{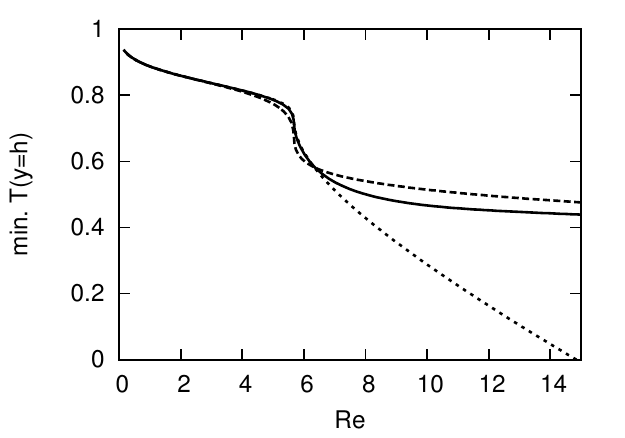}} \\
  \subfigure[$\PR\,=\,30$, $\BITILDE\, =\, 0.1$]{\label{fig-AutoC}
  \includegraphics[width=0.48\textwidth]{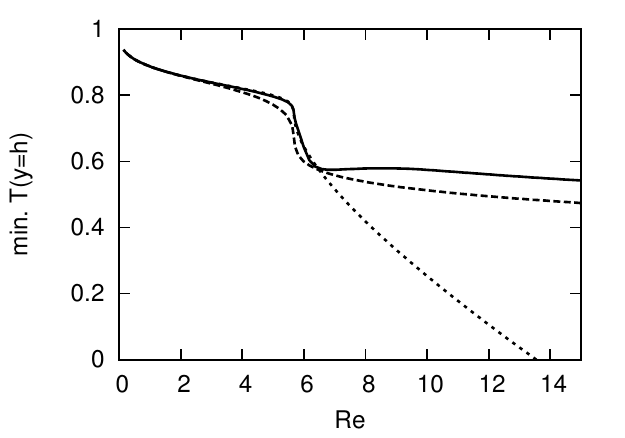}} \hfill
  \subfigure[$\PR\,=\,7$, $\BITILDE\, =\, 10$]{\label{fig-AutoD}
  \includegraphics[width=0.48\textwidth]{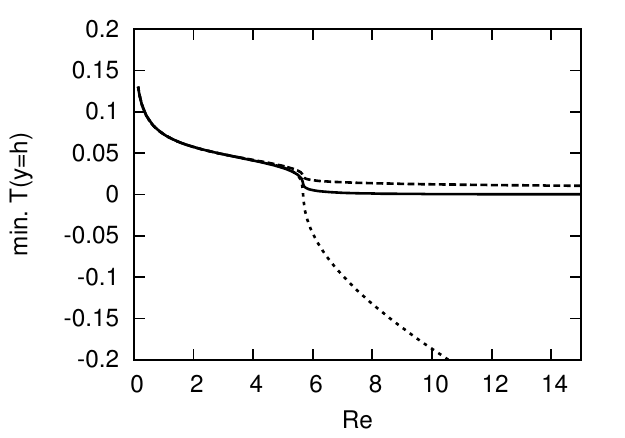}}
  \caption{\small\em Free surface temperature as function of the \textsc{Reynolds} number for solitary-wave solutions to the \textsc{Fourier} equation (solid lines), to system \eqref{eq:fah} (dashed line) and to the averaged energy balance \eqref{Ruy05} (dotted line).}
  \label{fig:Auto}
\end{figure}

We first consider the homoclinic orbits solutions to \eqref{eq:fah} for the set of parameters investigated by \cite{Trevelyan2007}, \ie $\PR\ =\ 7\,$, $\BITILDE\ =\ 0.1$ and $\KA\ =\ 30000$ (see Figure~11 in that reference). These values correspond well to the typical situation of a water film. Figure~\ref{fig-AutoB} compares the minimum of the temperature at the free surface $\left.T\,\right|_{z\, =\, h}$ obtained from \eqref{eq:fah} (dashed lines) with the solution to the \textsc{Fourier} equation (solid lines) and to the averaged energy balance \eqref{Ruy05} (dotted lines). The curves present a change of behaviour around $\RE\ =\ 6$ which signals the onset of capillary roll-waves in the drag-inertia regime identified by \textsc{Ooshida} \cite{Takeshi1999} where inertia effects become dominant (see \cite{Chakraborty2014}). Below this value of the \textsc{Reynolds} number, the curves are very close as a result of the consistency with the long-wave expansion up to first order. Discrepancies are observable whenever inertia effects are significant. They are emphasized when the \textsc{Prandtl} number is raised to $\PE\ =\ 30$ (compare Figures~\ref{fig-AutoB} and~\ref{fig-AutoD}). Note that all values of the free surface temperature corresponding to \eqref{eq:fah} fall into the admissible range $[0,\,1]\,$, whereas solutions \eqref{Ruy05} do not exhibit non-physical negative values of $\left.T\,\right|_{z\, =\, h}$ for Biot numbers larger than $0.1$. The proposed conservative formulation therefore corrects the main drawback observed with the weighted residual method yielding \eqref{Ruy05}.

\begin{figure}
  \centering
  \subfigure[wave profile]{\label{fig:Auto-2a}
  \includegraphics[width=0.48\textwidth]{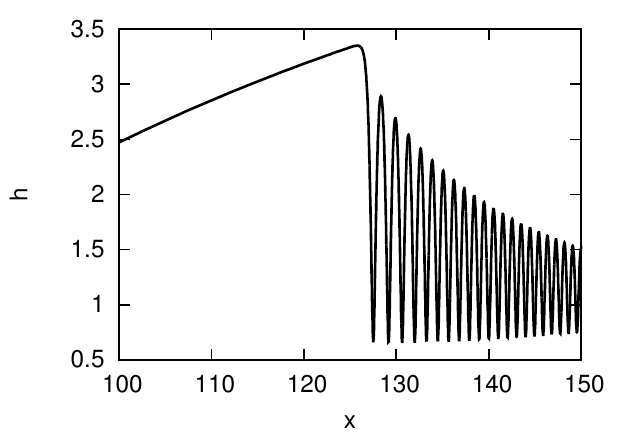}}\hfill%
  \subfigure[$\bt$]{\label{fig:Auto-2b}
  \includegraphics[width=0.48\textwidth]{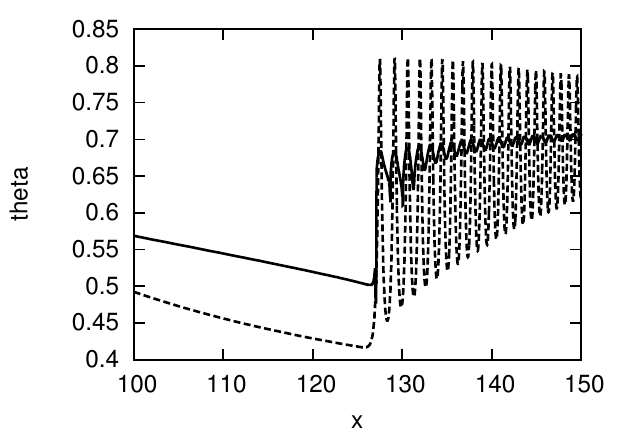}}\\
  \subfigure[$\bt$ vs. $h$ (Fourier)]{\label{fig:Auto-2c}
  \includegraphics[width=0.48\textwidth]{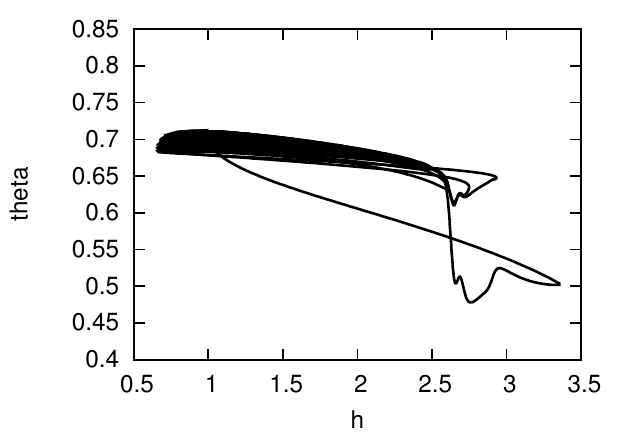}}\hfill%
  \subfigure[$\bt$ vs. $h$ (model) ]{\label{fig:Auto-2d}
  \includegraphics[width=0.48\textwidth]{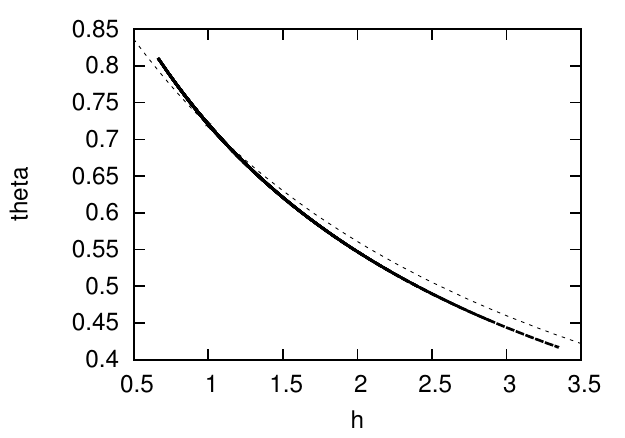}}
  \caption{\small\em Comparisons of the solutions to the \textsc{Fourier} equation and to the model for a large solitary wave. The solid (dashed) line refers to the \textsc{Fourier} (model) solution. $\bt\0(h)\ =\ 1/(1\,+\,\B\, h)$ is compared to $\bt$ in panel~(d) (thin dotted line).}
  \label{fig:Auto-2}
\end{figure}

Figure~\ref{fig:Auto-2} compares the solutions to the model and to the \textsc{Fourier} equation in the absence of a coupling between the hydrodynamics and the heat transfer ($\MA\ =\ 0$ and $\Pi_\mu\ =\ 0$). For both computations, the hydrodynamics is modelled by \eqref{ouf} assuming a parabolic velocity distribution. The parameter values correspond to the most demanding case tested, \ie $\PR\ =\ 30\,$, $\RE\ =\ 30$ and $\BITILDE\ =\ 0.1\,$. The wave profile, displayed in Figure~\ref{fig:Auto-2a}, presents an evident asymmetry  with a significant accumulation of capillary waves at its front. At a first glance, the distributions of temperature at the free surface seems similar for the \textsc{Fourier} equation and for the model (see Figure~\ref{fig:Auto-2b}). The model predicts correctly the amplitude of variation of the free-surface temperature but significantly overestimates the fluctuations of $\bt$ at the capillary region of the wave. Panels~(c) and (d) display the evolutions of $\bt$ as a function of the free-surface elevation $h$. This representation reveals important differences between the two solutions. The solution to the \textsc{Fourier} equation presents a sharp variation of $\bt$ at $h\ \approx\ 2.6\,$, which corresponds to the location of a stagnation point at the front of the wave in its moving frame ($u\,(y=h)\ \approx\ 3v/2\ =\ c$). This stagnation point marks the limit of extension of the recirculation region under the wave. At this point, cold liquid is dragged from the top of the waves, whereas hot liquid is pulled from the first trough region, which results into the formation of a thermal boundary layer and a sharp variation of $\bt$ \cite{Trevelyan2007}. The model does not reproduce this feature, the free-surface temperature $\bt$ following closely the evolution of the surface elevation $h$ as $\bt\ \approx\ \bt\0(h)$ (compare thin-dotted and dashed lines in panel~d).

The failure of the model to predict the onset of a thermal boundary layer at large values of $\PE$ stems from the representation choice of the convection process in \eqref{eq-bt-2} and from the expression of ${\tilde g}$, which artifially constrain $\bt$ to converge to $\bt\0(h)$ in the limit of large \textsc{P\'eclet} numbers.  We note that changing the convective terms in \eqref{eq-bt-2} to $h\partial_t\, g\ +\ (3q/2)\,\partial_x g$ as suggested by the above discussion does not cure this limitation as our tests show that $\bt$ remains close to $\bt\0(h)$ in that case. We propose to follow this line of thought in a fortcoming publication.

%%% ------------------------------------------------------------------------ %%%

\section{Conclusions \& Perspectives}
\label{sec:disc}

In this article, a new model for heated falling films is proposed: whereas the hydrodynamic part has already been known (see, for example, \cite{Noble2007}), the asymptotic model associated to the heat field is derived to preserve the conservative form of the averaged equations, which enables to make use of efficient numerical methods. The result corresponds to an explicit conservative model consistent up to the order $\O(\eps)$ with the \textsc{Fourier}--\textsc{Navier}--\textsc{Stokes} equations. The numerical experiments and comparisons with previous works (\eg \cite{Trevelyan2007}) and with 2D heat equation direct simulations validate the proposed model. The proposed model accounts for the coupling between hydrodynamics and heat transfer  by two physical mechanisms:
\begin{itemize}
  \item \textsc{Marangoni} effect: the surface tension coefficient is supposed to be \emph{linearly} dependent on the fluid temperature $T$
  \item Viscosity effect: the dynamic viscosity $\mu$ is also supposed to be \emph{linearly} dependent on the fluid temperature $T\,$.
\end{itemize}
To our knowledge these two effects have been considered separately, but their coupling seems to have been investigated only in \cite{DAlessio2014}. An important achievement of the proposed model is the absence of the non-physical negative values of the temperature that have been observed previously with the models proposed by \cite{Ruyer-Quil2005, Scheid2005, Trevelyan2007} using the weighted-residual method. Owing to an appropriate writing of the convective terms in the heat balance, the free-surface temperature $\bt$ remains in the physically admissible range of values in all considered cases. However, the price to be paid for this achievement is to artificially constrain the temperature field to follow the kinematics of the film in the high \textsc{P\'eclet} limit ($\bt\ \approx\ \bt\0$). To our opinion, it is unlikely that a modelling of the heat transfer with a unique averaged heat balance which relaxes the limitations stated above is achievable. This suggests to add more variables to represent the temperature field in order to account for the departures of the temperature distribution from the linear assumption \eqref{T-Nusselt}. We intend to follow this line of thought in a separate study.

%%% ------------------------------------------------------------------------ %%%

\subsection*{Acknowledgments}
\addcontentsline{toc}{subsection}{Acknowledgments}

The authors acknowledge a financial support from CNRS (INSIS, Cellule \'Energie, exploratory project call $2014$). Moreover, we wish to thank Didier~\textsc{Bresch} for helpful discussions on the topic of falling films modeling.

%%% ------------------------------------------------------------------------ %%%

%%% Bibliography
\addcontentsline{toc}{section}{References}
\bibliographystyle{abbrv}
\bibliography{biblio}

\end{document}